\documentclass[conference]{IEEEtran}
\usepackage{cite}
\usepackage{amsmath,amssymb,amsfonts}
\usepackage{amsthm}
\usepackage{algorithmic}
\usepackage{graphicx}
\usepackage{textcomp}
\usepackage{xcolor}
\usepackage[hyphens]{url}
\usepackage{svg}
\usepackage{tabulary}
\usepackage{caption} 
\usepackage{subcaption}
\usepackage{multicol}
\usepackage{multirow}
\usepackage{dblfloatfix}
\usepackage{float}
\usepackage{braket}
\usepackage{authblk}
\usepackage[normalem]{ulem}
\long\def\comment#1{}

\def\parah#1{\vspace*{0.0in} \noindent{\bf #1:}}

\def\BibTeX{{\rm B\kern-.05em{\sc i\kern-.025em b}\kern-.08em
    T\kern-.1667em\lower.7ex\hbox{E}\kern-.125emX}}

\title{Quantum Hardware Roofline: Evaluating the Impact of Gate Expressivity on Quantum Processor Design}
\author[1]{Justin Kalloor}
\author[1]{Mathias Weiden}
\author[2]{Ed Younis}
\author[1]{John Kubiatowicz}
\author[2]{Bert De Jong}
\author[2]{Costin Iancu}
\affil[1]{Department of Electrical Engineering and Computer Science, University of California, Berkeley}
\affil[ ]{\textit{\{jkalloor3, mtweiden, kubitron\}@cs.berkeley.edu}}
\affil[2]{Computational Research Division, Lawrence Berkeley National Laboratory}
\affil[ ]{\textit{\{edyounis, wadejong, cciancu\}@lbl.gov}}

\begin{document}
\maketitle
\thispagestyle{plain}
\pagestyle{plain}


\begin{abstract}
The design space of current quantum computers is expansive with no obvious winning solution.
This leaves practitioners with a clear
question: ``What is the optimal system configuration to run an algorithm?''. This
paper explores hardware design trade-offs across NISQ systems to guide algorithm and hardware design choices. The evaluation is driven by algorithmic workloads and algorithm fidelity models which capture architectural features such as gate expressivity, fidelity, and crosstalk. We also argue that the criteria for gate design and selection should be extended from maximizing average fidelity to a more comprehensive approach that takes into account the gate expressivity with respect to algorithmic structures. We consider native entangling gates (CNOT, ECR, CZ, ZZ, XX, Sycamore, $\sqrt{\text{iSWAP}}$), proposed gates (B Gate, $\sqrt[4]{\text{CNOT}}$, $\sqrt[8]{\text{CNOT}}$), as well as parameterized gates (FSim, XY). Our  methodology is driven by a custom synthesis driven circuit compilation workflow, which is able to produce minimal circuit representations for a given system configuration.
By providing a method to evaluate the suitability of algorithms for hardware platforms, this work 
emphasizes the importance of hardware-software co-design for
quantum computing.

\end{abstract}

\section{Introduction}
\label{section:intro}
Quantum computers offer an exciting opportunity to explore problems
previously considered intractable. An assortment of companies have introduced gate-based quantum machines that range in qubit technology:
superconducting transmon qubits \cite{barends_transmon},
fluxonium qubits \cite{bao_fluxonium_2022},
trapped-ion qubits \cite{cirac_ions}, neutral
atoms \cite{Jaksch_neutral}, and several others. 

While existing hardware offerings are not mature enough to provide
realistic quantum advantage~\cite{aaronson_complexity-theoretic_2016}
or quantum utility~\cite{kim_evidence_2023}, they can be used as a
good indicator of the future. Practical questions have
already arisen in the community related to the comparison of different
hardware solutions: ``What computer should I use to run my algorithm? How can I improve my current quantum processor? What  gates should
I provide to end-users?"

In the current Noisy Intermediate Scale Quantum (NISQ) computing era,
gates are imperfect and introduce error in program outputs.
Thus, the most important performance metric for current systems is their ability to execute algorithms with the least amount of error, i.e. maximize  algorithmic fidelity.

To this end, hardware designers attempt to improve the accuracy of an algorithm's execution using a multi-stage design process aimed at optimizing behavior across multiple hardware characterization
criteria: gate fidelity, crosstalk (gate parallelism and qubit
connectivity), etc. This process, centered around gate fidelity, proceeds as follows: First choose a native
entangling two-qubit gate that can be implemented with high fidelity,
and that is ``good enough'' to represent any two qubit process
(unitary). After this, develop additional techniques, e.g. crosstalk
mitigation, to further improve gate fidelity.

We believe that this design process can be improved upon from both a hardware and end-user perspective.  Most hardware characterization
metrics (\cite{magesan_robust_2011, cross_validating_2019, knill_randomized_2008}) are gate and algorithm agnostic; therefore, these widely accepted metrics (e.g. gate fidelity) are hard to correlate directly with algorithm performance across systems with distinct hardware characteristics. Full algorithm fidelity models that capture hardware characteristics (gate fidelity and parallelism/crosstalk) have been introduced in the literature \cite{chen_benchmarking_2023}. While they are able to assess the fidelity of an algorithm when executed on a single hardware configuration, these models still lack predictive power when varying architectural parameters. The problem stems from the fact that these metrics combine algorithm-agnostic hardware characterization metrics with metrics that characterize the program implementation and resource consumption (e.g. gate count, circuit depth), and implicitly
the impact of the program generators and compilers.

\comment{The main metric used to evaluate these questions is gate fidelity. In
general, NISQ architects try to provide gates that can easily
decompose from a standard gate set while maximizing fidelity. We
provide in Table \ref{tab:device_summary} the current set of gates
provided by major vendors and the gate fidelities provided. Other
metrics used to evaluate hardware include benchmark based protocols
such as Quantum Volume and Algorithmic
Qubits \cite{cross_validating_2019, chen_benchmarking_2023}. While
these do take a top-down approach to benchmarking hardware, they
provide no information on quantifying the trade off between a gate's
fidelity and ability to implement a circuit. \par
}


In this paper we argue that gate set design should be driven by representational power in the context of a given algorithm or algorithmic workload.
In order to attain the most resource efficient implementation, we use custom compilation workflows that combine traditional compilers, such as Cirq~\cite{Cirq} or Tket~\cite{sivarajah_tketrangle_2021}, with circuit synthesis tools. Note that the inferences made in this paper could not be
obtained without leveraging the BQSKit~\cite{bqskit} circuit synthesis tools.

The evaluation is driven by a workload that contains several
algorithms of wide interest, such as QFT, QAOA \cite{farhi_quantum_2014}, TFIM models \cite{shin_phonon-driven_2018}, Quantum Finance algorithms \cite{herman_survey_2022}, and Quantum Machine Learning models \cite{cerezo_challenges_2022}. For each algorithm we consider problem instances of increasing scale (qubit count) and generate the most resource efficient implementation for a given hardware configuration (gate set and qubit interconnect topology). We consider native entangling gates (CNOT, ECR, CZ, ZZ, XX, Sycamore, $\sqrt{\text{iSWAP}}$), proposed gates (B Gate, $\sqrt[4]{\text{CNOT}}$, $\sqrt[8]{\text{CNOT}}$), as well as parameterized gates (FSim, XY), together with several
qubit interconnect topologies. 

\begin{table*}
\centering
\begin{tabulary}{\textwidth}{|L|L|L|L|L|L|}
\hline
Hardware      & \# Qubits  & Technology   & Connectivity & Gate Set            & 1Q/2Q Error(RB/XEB) \\ \hline
Google Sycamore  & 54 & Superconducting & Mesh            & 1Q: XZ, RZ 2Q:FSim,$\sqrt{\text{iSWAP}}$, Sycamore, CZ & 0.001 / 0.01           \\ \hline
IBM Eagle r3    & 127     & Superconducting & Mesh            & 1Q: SX, RZ, X 2Q: ECR     & 0.0002 / 0.007          \\ \hline
IBM Hummingbird r3 & 65      & Superconducting & Mesh            & 1Q: RZ, SX, X 2Q: CNOT     & 0.00027/ 0.012          \\ \hline
IBM Falcon r5   & 27      & Superconducting & Mesh            & 1Q: RZ, SX, X 2Q: CNOT     & 0.0003/ 0.0079          \\ \hline
Rigetti Aspen M3  & 79      & Superconducting & Mesh             & 1Q: RX, RZ 2q: CZ, XY     & 0.001/ 0.14, 0.092        \\ \hline
Quantinuum H2   & 32      & Ion Trap    & A2A             & 1Q: U1, RZ 2Q: ZZ       & 3e-5/0.002            \\ \hline
IonQ Forte (2024) & 35      & Ion Trap    & A2A             & 1Q: GPi, 2Q: ZZ, XX      & 0.0002/ 0.0040          \\ \hline
IonQ Aria     & 21      & Ion Trap    & A2A             & 1Q: GPi, 2Q: XX        & 0.0006/ 0.0040          \\ \hline
\end{tabulary}
\caption{\it \footnotesize   Summary of existing commercial Quantum Computing hardware \cite{Qiskit, Cirq, rigetti_data, ionq_data, atom_computing_data}. As we can see, most devices available now are superconducting or ion trap devices, with superconducting devices proving to be easier to scale. This contrasts the ion trap devices which show on average higher RB fidelity. Additionally, superconducting qubits have a mesh (2D Nearest Neighbors) topology while ion traps are all-to-all (A2A).}
\label{tab:device_summary}
\end{table*}

This paper makes the following contributions:

First, we introduce analytical 
models that combine hardware (gate fidelity, parallelism and qubit
connectivity)  and
algorithm implementation (gate count, depth) characteristics to provide useful guidance for hardware and compiler designers, as well as system end-users. 
We introduce a comparative performance {\it roofline} based approach which is able to derive which particular metric can lead to overall improvements in algorithmic fidelity, as well as upper bounds on these metrics past which no additional end-user gains can be expected. For example, when comparing Sycamore (Google) and
CNOT (IBM) entangling gates for a particular algorithmic workload, our analyses show that there are ranges of relative gate fidelities where one configuration can always outperform the other. Once a certain {\it threshold} fidelity has been attained,
no improvements in one- or 
two-qubit gate fidelity on any architecture can lead to better relative performance with respect to the other.

Next, we introduce a circuit synthesis based compilation procedure which indicates that the existing gate set design criteria that favors choosing gates based on their attainable fidelity and representational power of random two-qubit process may be misleading. Instead, our analysis shows that the criteria should
be augmented with their representational power for multi-qubit processes
(e.g. three qubit) that are drawn from implementations of existing algorithms. For example, while the B-gate~\cite{zhang_minimum_2004} is the most expressive gate for two qubit unitaries, we cannot uncover advantages when using it to represent complex programs. When compared against CNOT, B-gates lead to gate count increases and possible fidelity decreases.
At the other end of the spectrum, we show that several
low-entanglement gates such as $\sqrt[4]{\text{CNOT}}$ and $\sqrt{\text{iSWAP}}$ are sometime able to offer similar expressive performance as maximally entangling gates for important circuits such as TFIM, QFT, and QAE, leading to better algorithmic fidelity.

As discussed in Section~\ref{sec:disc}, we believe our assessment procedure extends well beyond NISQ into the Fault Tolerant era of quantum computing.

\comment{
The rest of the paper is organized into the following sections:

\begin{enumerate}
    \item Section \ref{section:arch} will provide a brief overview of the quantum machines that exist today, as well as current comparison metrics and why they fail.
    \item Section~\ref{section:process} gives an overview of our benchmarking process.
    \item In Section \ref{section:fid_model} we provide a detailed explanation of our fidelity models and subsequent analysis on the importance of single qubit gates.
    \item  Section \ref{section:evaluation} presents the results of our process applied to a wide set of algorithms, organized by circuit class, to several different native gate sets. We then further explore the impact of topology on our models.
    \item Section \ref{section:discussion} discusses what trends are seen and offers insight into the behavior by analyzing them through the lens of the Weyl Chamber. We also show the behavior of additional architectural features, such as providing multiple entangling gates and parameterized entangling gates.
    \item Section \ref{section:conclusion} summarizes our findings and presents future directions to be explored in the field of software hardware co-design.
\end{enumerate}
}

\section{Quantum Hardware Characterization and Benchmarking}
\label{section:arch}

Today's systems are dominated by superconducting (IBM, Google) and
trapped ion (Quantinuum, IonQ) qubits. Neutral Atom (QuEra, Atom
Computing) and silicon-spin qubits (Intel) are starting to gain
traction , and several other technologies are being developed. All these systems
expose to end-users a universal gate set~\cite{Nielsen_2002}, composed of single-qubit and entangling two-qubit native gates. These gate-set choices are outlined in
Table \ref{tab:device_summary}. Several methods exist to characterize today's quantum
machines:


\comment{ 
\begin{figure}[ht]
  \centering
  \includegraphics[width=.3\textwidth]{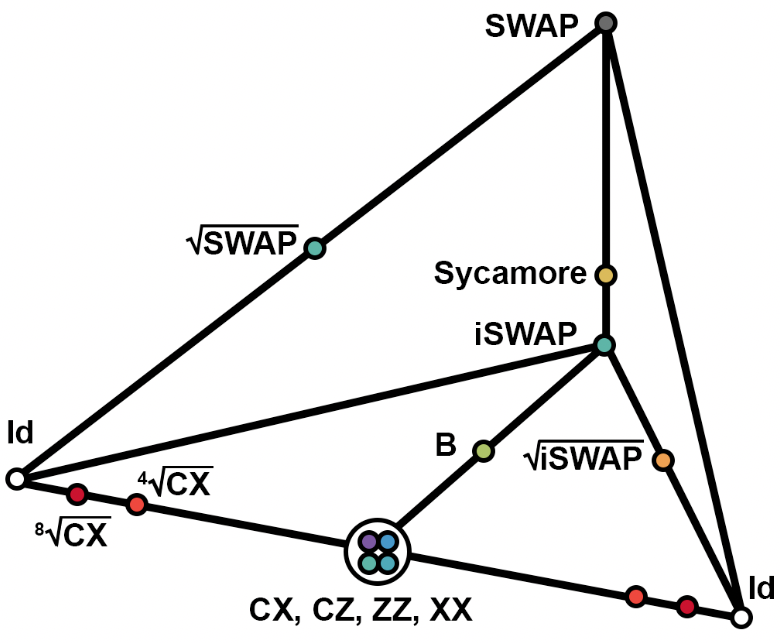}
  \caption{Projection of various 2-qubit gates onto the Weyl Chamber. Gates that are in the same point represent unitaries that differ by single qubit rotations applied to each qubit (a 1:1 mapping).}
  \label{fig:weyl_chamber}
\end{figure}
}

\parah{Average Gate Fidelity} The widest
used characterization and processor optimization metric is the {\it average gate fidelity}, which captures the probability that
a state does not succumb to any error when a gate is applied.
Fidelity can be measured using Randomized Benchmarking (RB) protocols~\cite{magesan_characterizing_2012, magesan_robust_2011, knill_randomized_2008}, which
use random Clifford gates to create an identity channel over a set of
qubits. Averaged over a randomized group of these Clifford circuits,
the error of the quantum channel simplifies to a depolarization
channel with a single probability $p$, the average infidelity of the gate's application. By using variable
lengths of random Clifford circuits, existing protocols
calculate the infidelity per gate ($p_{\text{gate}}$): fidelity is then computed as $1 -p_{\text{gate}}$.

In 2019, Google introduced the cross
entropy benchmarking (XEB) protocol \cite{arute_supplementary_2019} as another way to compute average
gate fidelity. Importantly, this study also shows that the total
average fidelity of a circuit can be approximated using a simple
{\it digital error model}, validated for NISQ size systems: \[ \mathbf{F_d}
= \prod_{i=1,2} f_i^{n_i} \] where $n_i$ is the number of $i$-qubit
gates in the circuit, and $f_i$ is the average fidelity of an
$i$-qubit gate. 

As RB protocols have trouble scaling past three qubit processes, Cycle
Benchmarking (CB) \cite{Erhard_2019} has been proposed to improve
characterization scalability to larger processes (and hardware). CB
based protocols indicate that besides average gate fidelity, hardware
dependent metrics such as qubit connectivity and algorithm specific
metrics such as gate parallelism per cycle need to be taken into
account when assessing algorithm fidelity.

\comment{better
describe the error channel of a quantum device. Gates of interests
could be benchmarked by adding in random Pauli cycles. These Pauli
cycles collapse the quantum error channel to a Pauli Error Channel,
which can then be benchmarked using a similar process to Randomized
and Cross-Entropy Benchmarking. This benchmarking is able to provide
an error per Pauli Operation instead of a single error for the entire
channel. }

\parah{Quantum Volume} Quantum Volume (QV) \cite{cross_validating_2019} characterizes the capability of hardware to 
execute random circuits of a certain size. This metric cannot be used to compare the fidelity of different process implementations running on the same machine or that of a single process running across different machines.

\parah{Algorithmic Qubits} IonQ's Algorithmic Qubits (AQ) metric ~\cite{chen_benchmarking_2023} 
captures a system's ability to execute an algorithmic workload. AQ
protocols measure the largest number of effectively perfect qubits you
can deploy for a typical quantum program. It is similar to QV, but it
additionally considers quantum error correction and presents a clear
and direct relationship to qubit count. The AQ metric captures
the impact of the compilation tool-chain.


All of these protocols and metrics reveal different useful information
about a single configuration of quantum machine. However, they all fail when
comparing across different hardware and gate sets. While we can
measure the fidelity and quantum volume of a CNOT-based machine and of a
Sycamore-based machine, this does not give us any information on their
respective abilities to run a given algorithm. AQ encapsulates the algorithmic potential of different
hardware, but it is still unable to quantify the degree to which one needs make
changes to an architecture's configuration in order to provide better comparative
performance.

In order to make these inferences, we advocate for an algorithmic
workload based approach centered around circuit fidelity
models that combine hardware characterization metrics with the
hardware's ability to represent and implement a particular algorithm. We start with the simple digital model based on average gate
fidelity, which we then extend to account for crosstalk due to
parallel gate execution as well as qubit connectivity (an idle qubit can be affected when executing an operation on a neighboring
qubits).

\section{Quantum Algorithms and Compilers}
\label{section:process}
The digital model indicates that circuit fidelity is determined by the
average gate fidelity and the circuit gate count: improving both
metrics will improve algorithmic fidelity. The gate count for a given
algorithm implementation is determined by hardware characteristics: 1)
representational power of the native gate set; and 2) qubit
interconnection topology.

Due to exponentially compounding gate infidelities, the dominant factor in the digital model is gate count. This has two consequences: 1) comparisons between system configurations should be done using the implementation with the fewest number of gates attainable, together with the lowest depth or highest gate parallelism; and 2) the compilation
tool-chain plays a very important role in determining the overall
``performance'' of a given configuration.

\subsection{Quantum Algorithms}

Domain generators~\cite{openfermion, Qiskit} that produce the circuit associated with a given algorithm tend to have the following common characteristics:
\begin{itemize}
\item They are developed to generate circuits in a restricted gate set. Most generators use directly the CNOT gate, while some hardware-vendor-provided generators target only vendor supported native gates. Thus, the compiler's ability to translate a circuit between gate sets is paramount for accurate architectural comparisons.
\item The generated circuits have a logical qubit connectivity that resembles the domain level structure. For example, optimal QFT circuits are generated assuming an all-to-all qubit connectivity. Circuits generated for fermionic \cite{openfermion} interactions map fermions to qubits using a logical topology that resembles the structure of the physical system modeled. Thus, the quality of the routing algorithm within compilers matters for accurate architectural comparisons.
\item Some generators are deemed as optimal. Optimality here relates to asymptotic complexity: a good compiler can greatly reduce the constants that appear in the complexity formula. Again, a good quality compiler is paramount. 
\end{itemize}

\subsection{Quantum Compilers}

\begin{figure}
  \centering \includegraphics[width=0.42\textwidth]{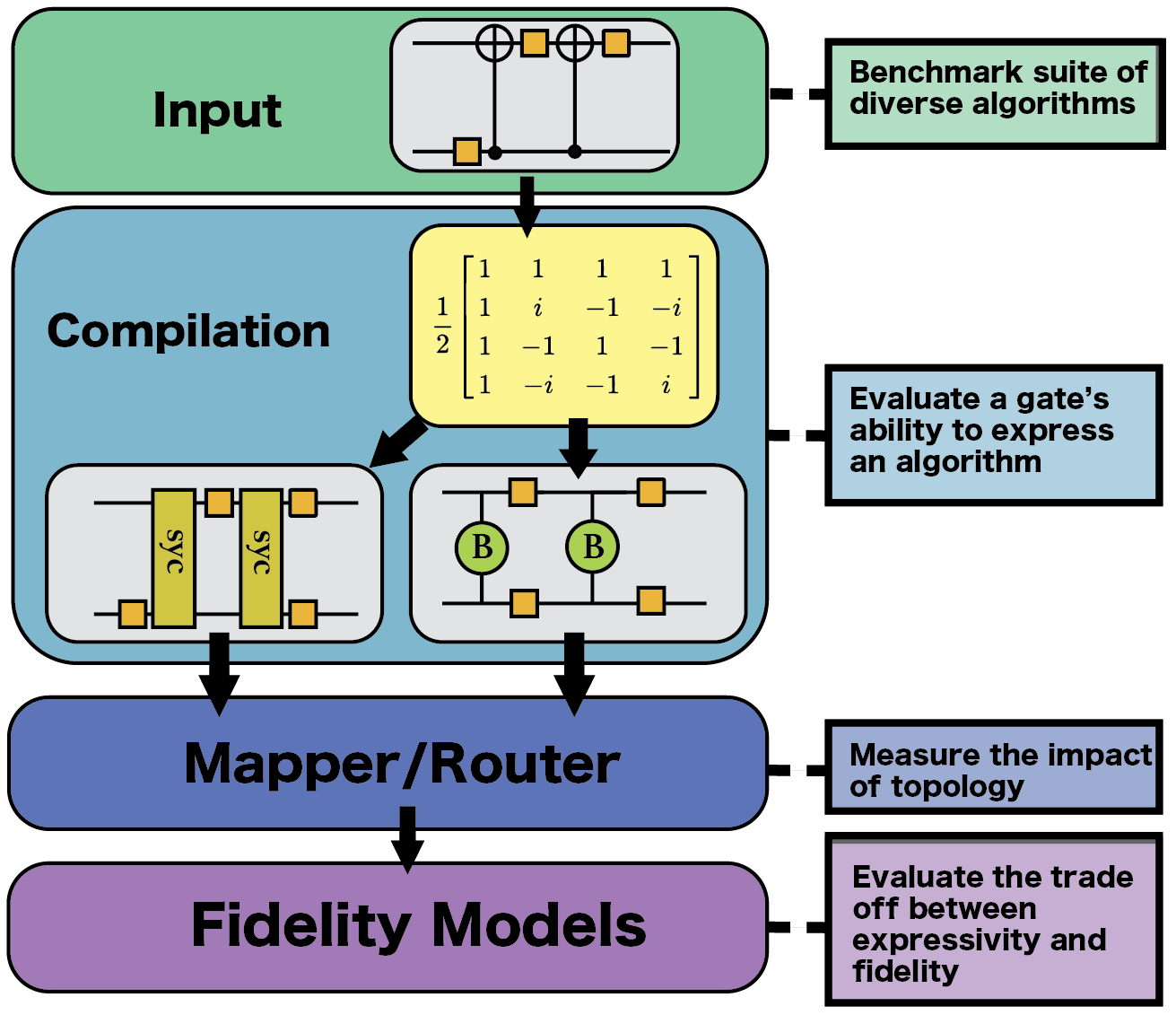} 
  \vspace*{-.05in}
  \caption{\footnotesize \it Our Hardware Comparison Procedure:  A synthesis-based cross-compilation process (sometimes called ``transpilation'') allows us to explore multiple gate sets and there ability to express an algorithm in terms of gate count, depth, and parallelism. From there, we can understand the effects of topology
  and fidelity on the overall performance of a quantum
  machine.} \label{fig:process}
\end{figure}

Conceptually, quantum compilers perform several circuit transformation functions: 1) eliminate
gates that are redundant or can be simplified from the circuit; 2)
map and route the input circuit to the hardware configuration; and 3)
translate (transpile) the input circuit to a different gate set.  As seen in Figure~\ref{fig:process}, compilation forms an important part of our evaluation process.

Traditional vendor compilers use peephole optimizations based on
2-qubit gate synthesis (KAK~\cite{tucci2005introduction} decomposition), application of gate commutativity rules, or domain specific pattern rewriting rules (\emph{e.g.} Tket's phase gadgets~\cite{Cowtan_2020}). They also provide mapping and routing algorithms~\cite{Qiskit,sivarajah_tketrangle_2021, li2019tackling} and
compilation to multiple gate sets (``transpilation''). In particular, transpilation is performed using a one-to-one gate rewriting rule: any two-qubit gate is rewritten directly from one (e.g. CNOT) to another (e.g. Syc).

\begin{table*}
\begin{minipage}{0.34\textwidth}
\centering
\begin{tabular}{l|p{0.75\linewidth}}
\textbf{Family} & \textbf{Benchmarks (circuit\_width)}  \\ \hline
TFIM   & TFIM\_16, TFIM\_64, TFXY\_16, TFXY\_64 \\
QAE       & qae\_9, qae\_13, qae\_17, qae\_21   \\
QFT       & qft\_4, qft\_12, qft\_64        \\
QAOA       & qaoa\_10                \\
QPE       & qpe\_14, qpe\_18            \\
Adder      & adder\_9, adder\_63, mul\_10, mult\_60 \\
Shor       & shor\_12, shor\_16, shor\_24, shor\_28 \\
Grover      & grover\_5      \\
Hubbard     & hubbard\_4, hubbard\_8, hubbard\_12   \\
QML       & qml\_6, qml\_13, qml\_22        \\ 
VQA       & vqe\_12 (LiH), vqe\_14 ($\text{BeH}_2$) \\
\end{tabular}
\caption{\footnotesize \it Our Benchmarks: List of benchmark circuits organized by family. Each circuit was initially created with CNOT and U3 gates.}
\label{tab:benchmarks}
\end{minipage}%
\hfill
\begin{minipage}{0.63\textwidth}
\centering
\begin{tabular}{|c|c|c|cc|cc|cc|cc|}
\hline
\textbf{Benchmark}         & \textbf{Size}  & \textbf{Gate} & \multicolumn{2}{c|}{\textbf{BQSKit}}       & \multicolumn{2}{c|}{\textbf{Tket}} & \multicolumn{2}{c|}{\textbf{Cirq}} & \multicolumn{2}{c|}{\textbf{Qiskit}} \\ \hline
\multirow{4}{*}{\textbf{mul\_10}} & \multirow{4}{*}{163} &        & \multicolumn{1}{c|}{a2a}     & mesh     & \multicolumn{1}{c|}{a2a}  & mesh & \multicolumn{1}{c|}{a2a}  & mesh & \multicolumn{1}{c|}{a2a}  & mesh  \\ \cline{3-11} 
                  &           & cz      & \multicolumn{1}{c|}{\textbf{67}} & \textbf{89} & \multicolumn{1}{c|}{104}  & 134  & \multicolumn{1}{c|}{134}  & 227  & \multicolumn{1}{c|}{132}  & 141  \\ \cline{3-11} 
                  &           & b       & \multicolumn{1}{c|}{\textbf{110}} & \textbf{125} & \multicolumn{1}{c|}{208}  & 480  & \multicolumn{1}{c|}{-}   & -   & \multicolumn{1}{c|}{-}   & -   \\ \cline{3-11} 
                  &           & syc      & \multicolumn{1}{c|}{\textbf{103}} & \textbf{139} & \multicolumn{1}{c|}{208}  & 319  & \multicolumn{1}{c|}{260}  & 364  & \multicolumn{1}{c|}{-}   & -   \\ \hline
\multirow{3}{*}{\textbf{qft\_16}} & \multirow{3}{*}{264} & cz      & \multicolumn{1}{c|}{\textbf{237}} & \textbf{336} & \multicolumn{1}{c|}{240}  & 522  & \multicolumn{1}{c|}{264}  & 563  & \multicolumn{1}{c|}{264}  & 426  \\ \cline{3-11} 
                  &           & b       & \multicolumn{1}{c|}{\textbf{242}} & \textbf{302} & \multicolumn{1}{c|}{480}  & 1044 & \multicolumn{1}{c|}{-}   & -   & \multicolumn{1}{c|}{-}   & -   \\ \cline{3-11} 
                  &           & syc      & \multicolumn{1}{c|}{\textbf{241}} & \textbf{365} & \multicolumn{1}{c|}{480}  & 828  & \multicolumn{1}{c|}{288}  & 755  & \multicolumn{1}{c|}{-}   & -   \\ \hline
\multirow{3}{*}{\textbf{TFIM\_16}} & \multirow{3}{*}{240} & cz      & \multicolumn{1}{c|}{\textbf{200}} & \textbf{200} & \multicolumn{1}{c|}{240}  & 240  & \multicolumn{1}{c|}{240}  & 240  & \multicolumn{1}{c|}{240}  & 240  \\ \cline{3-11} 
                  &           & b       & \multicolumn{1}{c|}{\textbf{200}} & \textbf{202} & \multicolumn{1}{c|}{480}  & 480  & \multicolumn{1}{c|}{-}   & -   & \multicolumn{1}{c|}{-}   & -   \\ \cline{3-11} 
                  &           & syc      & \multicolumn{1}{c|}{\textbf{200}} & \textbf{208} & \multicolumn{1}{c|}{480}  & 219  & \multicolumn{1}{c|}{440}  & 440  & \multicolumn{1}{c|}{-}   & -   \\ \hline
\end{tabular}
\caption{ \footnotesize \it Comparison of two qubit gate counts for a subset of the benchmarks and gate sets across our different compilers. The first number under each compiler is for an all-to-all topology system, while the second number is for a mesh topology. Synthesis based compilers, such as BQSKit, produce the circuits with the least amount of gates. Note that Cirq (Qiskit) is unable to compile to the B Gate (B and Sycamore Gate).}

\label{tab:compiler_comp}
\end{minipage}
\end{table*}

Circuit synthesis~\cite{bqskit,squander} based tools have been
introduced recently and have been shown to provide better quality
implementations (~\cite{davis_qfast,liu_tackling_2023, younis_quantum_2022, qseed}) when compared against vendor compilers, albeit at the expense of increased compilation time. Some of these
tools integrate optimization~\cite{davis_qfast} with mapping~\cite{liu_tackling_2023} and gate transpilation~\cite{younis_quantum_2022}. They can 
search a large space of circuit structures and transformations. When considering transformations, these tools are able to perform global optimization of multi-qubit circuits, as opposed to peephole pattern replacement and one-one gate translation offered by vendor compilers. 

All compilation tools have one thing in common: the compilation workflow is custom and it consists of repeated applications of passes and transformations. While it is hard to quantify the impact of optimization, mapping, and transpilation phases in isolation, we note that compilers can realize up to an order of magnitude in gate count reduction, even when the input circuit is ``optimally'' generated.

\comment{\subsection{Mapping and Routing}
After transpilation, our circuits are then mapped onto a given topology. We run across multiple routers and mappers (\cite{sivarajah_tketrangle_2021, Qiskit, Cirq, wille_mqt_2023})in order to derive the best circuits (we show numbers from select algorithms in Table \ref{tab:compiler_comp}). In general, the PAM algorithm performs best by utilizing numerical synthesis at a block level and additionally accounting for permutations of these blocks to avoid adding additional SWAPs \cite{liu_tackling_2023}. For the PAM algorithm, we are able to upper bound the error (distance between our implemented unitary and target unitary) across the entire circuit by adding the error over all of the partitioned blocks in the circuit. We set this error upper bound to be $10^{-8}$ for each block in our circuit.}

\section{Evaluation Procedure}

We used the algorithms shown in Table
\ref{tab:benchmarks} for our evaluation. They include many
important categories including Variational Algorithms (VQA and QAOA)
\cite{peruzzo_variational_2014, farhi_quantum_2014},
Finance (QAE) \cite{herman_survey_2022}, Number Theory (QFT, QPE,
Shor) \cite{beauregard_circuit_2003}, Physical Simulation (Hubbard, Ising(TFIM))
\cite{bravyi_fermionic_2002, shin_phonon-driven_2018}, Search (Grover
\cite{mandviwalla_implementing_2018}), and Quantum Machine Learning
(QML). The QML circuit is based on 
\cite{cerezo_challenges_2022} and has an n-bit encoder and a two-local network. Most benchmarks were generated using Qiskit circuit generators\cite{Qiskit}, while the TFIM circuits were generated with F3C++ compiler \cite{Bassman_Oftelie_2022}. All input circuits were generated using CNOT entangling gates, which is standard practice. For each algorithm we generate several instances
across inputs and circuit sizes (number of qubits). Overall, we believe that our algorithmic workload provides a good sampling of the space of circuit implementations. We consider up to 64 qubit programs, with gate counts as high as 37000 accounting for a maximum depth of 44500. The logical topology of these programs ranges from linear in TFIM to the all-to-all connectivity in QFT.

\begin{figure*}[htbp!]
  \centering \includegraphics[width=\textwidth]{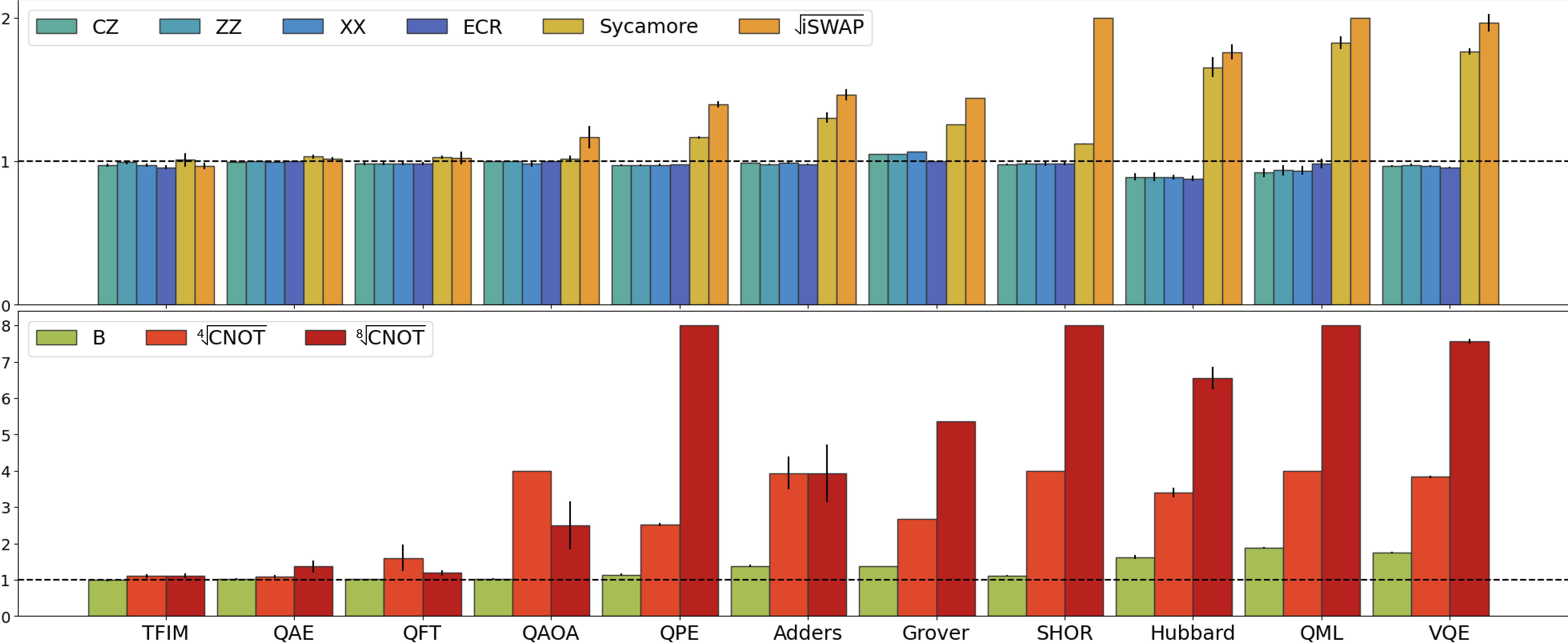}
  \caption{\label{fig:all_counts} \footnotesize \it
  Normalized two-qubit gate count for existing hardware (top) and
  theorized hardware (bottom). We plot the relative count with
  respect to the best attainable CNOT based circuit (best
  optimization across compilers). This shows the gate 
  count when using the logical algorithm connectivity, which correlates to the gate's respective ability to represent the algorithm.} 
\end{figure*}

We translate and optimize the benchmarks for  native gates
present in today's hardware (CNOT, ECR, CZ, ZZ, XX, Sycamore, $\sqrt{\text{iSWAP}}$), as shown in Table
\ref{tab:device_summary}. Additionally, we examine experimental gates theorized to provide algorithmic fidelity advantages
due to either high expressivity or high fidelity, B and
$\sqrt[4]{\text{CNOT}}$ or $\sqrt[8]{\text{CNOT}}$ respectively.

Figure \ref{fig:process} illustrates our process.  
We use a custom compilation workflow that composes compilers (Qiskit, Cirq, Tket) with the BQSKit~\cite{bqskit} circuit synthesis tools and selects the best circuits that result.  
Table \ref{tab:compiler_comp} shows a sampling of these results.  While details of the
compilation workflow are beyond the scope of this paper, to our
knowledge, we generate the best attainable implementations of a given
algorithm on a given hardware configuration. These resource
``optimal'' circuits\footnote{Meaning they do not improve with further compilation and optimization.}  are then
run through our fidelity models described in Section~\ref{section:fid_model} in order to compare overall performance.

\section{Gate Representational Power}
\label{section:expressivity}

Given a reference  implementation using CNOT gates, in
 Figure \ref{fig:all_counts} we show the relative two-qubit gate count
 (averaged across circuit families) after re-targeting to a particular
 gate. This data shows the ability of a gate to represent a particular algorithm, i.e. it captures its expressivity and  entanglement power.
When considering existing native gates, the \{CZ, ZZ, XX, ECR\} set seems to have the same representational power as CNOT gates.
The native gates \{Sycamore, $\sqrt{\text{iSWAP}}$\} have lower representational power than CNOT, as illustrated by their higher gate counts. When considering theorized gates, we see that \{B, $\sqrt[4]{\text{CNOT}}, \sqrt[8]{\text{CNOT}}$\} have overall lower representational power than CNOT. This is surprising as the motivation behind the introduction of the B-gate was its higher-than-CNOT representational power.

This behavior is also algorithm dependent. The Ising Model, QAE and QFT
circuits require almost the same number of gates, irrespective
of which gate it is used. For the rest of the circuits, we see more
nuanced behavior. The gate set \{CZ, ZZ, XX, ECR, CNOT\} leads to the
least amount of gates used, with much higher gate counts for the
set \{Sycamore, $\sqrt{\text{iSWAP}}$, B, $\sqrt[4]{\text{CNOT}}$, $\sqrt[8]{\text{CNOT}}$ \}.

Assuming that gates have different fidelities, the data indicate that
the machine with the highest fidelity will be best suited to execute TFIM,
QFT, and QAE. For the rest of the algorithms, suitability needs to be
examined while considering both gate fidelity and circuit structure (gate
count etc.). Our data also indicates that gate representational power with respect to full algorithms needs to be taken into account when selecting a system configuration. We discuss in detail representational power trade-offs in Section~\ref{section:discussion}.


%

\section{Circuit Fidelity Models}
\label{section:fid_model}

In the NISQ era, it is critical to maximize the probability that a circuit's output state is correct. The output expectation can be described a function of the average gate fidelity of the machine, written as \cite{Nielsen_2002}:
\[ \mathbf{F}_{\text{gate}} (\mathcal{E}, U)  = \int d\psi \bra{\psi}U^\dag \mathcal{E}(\psi)U\ket{\psi} \]
where U is the target unitary and $\mathcal{E}$ is the erroneous channel trying to implement U.

To assess algorithm fidelity on a particular system configuration, we use a series of models that capture circuit characteristics together with an increasing number of architectural features: (1) gate fidelity; (2) gate fidelity and parallelism. In Section \ref{sec:top} we discuss a model based on qubit connectivity as well.

Let $\mathbf{F}(\cdot)$ denote a circuit fidelity model, and let $A$ and $B$ denote two distinct system configurations. In order to enable system comparisons, we analyze the {\it objective function} given by:\centerline{ $\mathbf{\pi} = \mathbf{F}^A(\cdot) - \mathbf{F}^B(\cdot) $}

\subsection{Gate Fidelity} Our first model is derived from \cite{arute_supplementary_2019}, in which the authors verify that the measured fidelity and estimated fidelity based on this model track almost exactly for their tested circuits.

\newtheorem{definition}{Definition}
\begin{definition}[{\bf Digital Fidelity Model}] The average circuit fidelity $\mathbf{F}_d$ can be estimated as  
    \[ \mathbf{F_d} = \prod_{i=1,2,..} f_i^{n_i} \]
\end{definition}
    \noindent where $n_i$ is the number of $i$-qubit gates in the circuit, and $f_i$ is the average fidelity of an $i$-qubit gate. For systems with only one- and two-qubit native gates this becomes:
    \[ \mathbf{F_d}  = f_1 ^ {n_1} \cdot f_2^ {n_2} \]
with the objective function: 
        \[ \mathbf{\pi_d} = {f^A_1} ^ {n^A_1} \cdot {f^{A}_2}^ {n^A_2} - {f^B_1} ^ {n^B_1} \cdot {f^{B}_2}^ {n^B_2} \]

\subsection{Gate Fidelity and Parallelism}
Parallel execution impacts (negatively) the attainable gate average fidelity. To capture this, we use a model based on Cycle Benchmarking~\cite{Erhard_2019}. The protocol considers circuits as a series of cycles, with which we can calculate a single cycle fidelity as function of the 1-qubit and 2-qubit process fidelities
($\gamma$). Note that the process fidelity and average fidelity as
defined above are related by the following
equation \cite{horodecki_general_1999}:

\[
    f = \frac{d \cdot \gamma + 1}{d + 1}
\]

\noindent where $d = 2 ^ n$ is the dimension of the qubit register with $n$ qubits. We consider the register size of each cycle as the gate parallelism per circuit cycle. We average over the circuit parallelism and compute an average circuit register size.

\begin{definition}[{\bf Cyclic Fidelity Model}]

    \[\mathbf{F_c}  = \prod (1 - e_i * P_i)^{m} \]

\end{definition}

\noindent where $P_i$ is the average parallelism of $i$-qubit gates in the circuit, and $e_i$ is the average process infidelity for an $i$ qubit gate ($1 - \gamma_i$). $e_i$ can be measured using the Cycle Benchmarking protocol. The objective function for machine comparison becomes:
\vspace{-2mm}
\begin{align*}
    \mathbf{\pi_c} &= \mathbf{F_c}^A - \mathbf{F_c} ^B \\
    &= (1 - e^A_1 * P^A_1)^{m^A} \cdot (1 - e^A_2* P^A_2)^{m^A} \\
    &= (1 - e^B_1 * P^B_1)^{m^B} \cdot (1 - e^B_2* P^B_2)^{m^B} 
\end{align*}



\begin{figure}[h]
	\centering
	\includegraphics[width=0.45\textwidth]{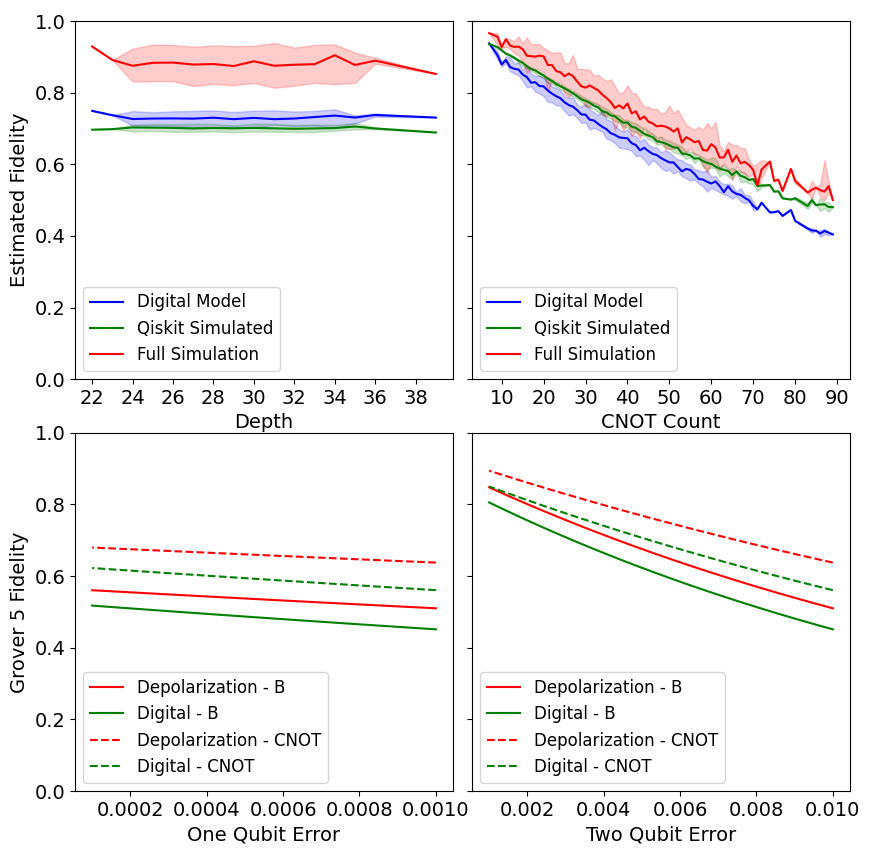}
    \vspace*{-0.1in}
	\caption{\footnotesize \it Validation experiments for our fidelity model. (a) Plot against full simulation on IBM Noisy Simulator and derived depolarization channel. We keep constant depth circuits with varied CNOT count gates. (b) Fidelity plot of constant count circuits as we vary circuit depth. (c) Varying the 1-qubit gate error of our model vs. a general depolarization model. (d) Varying the 2-qubit gate error of our model vs. a general depolarization model.
 }
	\label{fig:fig_valid}
\end{figure}

\subsection{Validation}

These models trade off accuracy for tractability. While full noise simulation is able to give the most accurate view of algorithmic fidelity, it does not scale to system sizes of interest. However, we can use it as a point of comparison to validate our models. We focus on validating the digital model for our experiments since that model is directly based on RB/XEB protocols which are available across all of our tested machines. \par
We first compare our algorithm on random circuits with varied gate count and depth (see Figures \ref{fig:fig_valid}a, \ref{fig:fig_valid}b). We compare our digital model against two other models: full simulation and a depolarization channel model. We use Qiskit's provided noisy backends which account for T1 and T2 coherence times of qubits, as well as explicit error channels for each gate application \cite{Qiskit}. We then run the Randomized Benchmarking protocol on this simulator to derive one- and two-qubit fidelities that we use to characterize a depolarization error channel. We expect our model to lower bound both other procedures, which we see in Figures~\ref{fig:fig_valid}a and \ref{fig:fig_valid}b. Importantly, the trends in all 3 lines are identical as we vary CNOT count and depth. \par
Secondly, we ensure that our model outputs similar relative results as we vary one- and 2-qubit error for different gate sets. We show these experiments in Figures \ref{fig:fig_valid}c and \ref{fig:fig_valid}d for the \emph{grover\_5} circuit transpiled to a CNOT machine and a B-gate machine. We see that the conclusions we would draw from either model are the same as we vary the error. We feel that the correlations seen in our model across these experiments in addition to the correlation seen in experiments run by Google on real hardware \cite{arute_supplementary_2019} validate the utility of our fidelity model.



\par

\section{Evaluating Quantum Machines}
\label{section:evaluation}

Now, we can finally answer the question  "{\it What machine should I use
to run my algorithm?}".

The answer is rendered trivial when considering currently published gate fidelity figures. 
Quantinuum boasts by far the highest 2-qubit gate fidelity at 0.998 and its ZZ
gate can express our algorithmic workload well. The H2 systems also provide
all-to-all connectivity which requires no additional routing. For algorithms 
that use more qubits than H2's capacity, the models suggest the IBM Eagle system.

A more interesting question is how could system configurations be changed in order to improve competitiveness, e.g. : \textit{``How can other machines become better than H2?''}.

\subsection{Quantifying Design Trade-offs}

Our procedure allows us to quantify the trade-offs between a gate's representational power for an algorithm and its fidelity.
This is a comparative analysis where we vary the models' parameters and solve for the objective function as defined in Section \ref{section:fid_model}. As gates continue to improve and calibration/noise-mitigation techniques advance, architects and end-users must consider:

\begin{enumerate}
    \item ``How much should I improve my gate fidelity in order to out-perform other machines? How does this vary by algorithm class?"
    \item ``Does single-qubit gate count matter for relative performance? At what point do we no longer care?"
    \item ``Does offering multiple entangling gates help with my machine's ability to express a circuit? Do parameterized entangling gates such as the FSim or XY gate help? If calibrating a more flexible machine leads to a drop in fidelity, how much drop can we afford?"
    \item ``How much of a fidelity improvement do I need to provide in order to overcome topological features of my machine?"
\end{enumerate}

Given that the maximum attainable gate fidelity is $1$, in order to compare NISQ-era devices, we want to use realistic constraints. First, to simplify the model, we will initially limit the single-qubit gate type to only the $U3$ gate. Current machines have parameterizable rotation gates that can be composed to perform any arbitrary single-qubit unitary. As 2-qubit gate errors still dominate, this simplification is justified. Based on
Table \ref{tab:device_summary}, single-qubit gate fidelities vary from around $0.999$ to $0.99999$ across all quantum machines
considered. For 2-qubit gates we range from fidelities of $0.990$ to $0.999$.


\subsection{Two-Qubit Gates Analysis}

We use the \emph{adder\_9} algorithm to directly compare IBM Falcon (CNOT) with Google Sycamore (Sycamore) machines as our driving example. The resulting objective function (defined in Section \ref{section:fid_model}) becomes:

\[\mathbf{\pi_d} = {f^A_1} ^ {70} \cdot {f^{A}_2}^ {49} - {f^B_1} ^ {91} \cdot {f^{B}_2}^ {66} \]

For comparisons, we rewrite the objective function to use the
Sycamore fidelity relative to CNOT. We vary the CNOT fidelity along the x-axis and the relative Sycamore fidelity along the y-axis. We then have two remaining free variables:

\[\mathbf{\pi_d} = {f^A_1} ^ {70} \cdot {x}^ {49} - {f^B_1} ^ {91} \cdot ({x \cdot y})^ {66} \]

\begin{figure*}
\begin{minipage}{\columnwidth}
    \centering
    \includegraphics[keepaspectratio=true,scale=0.52]{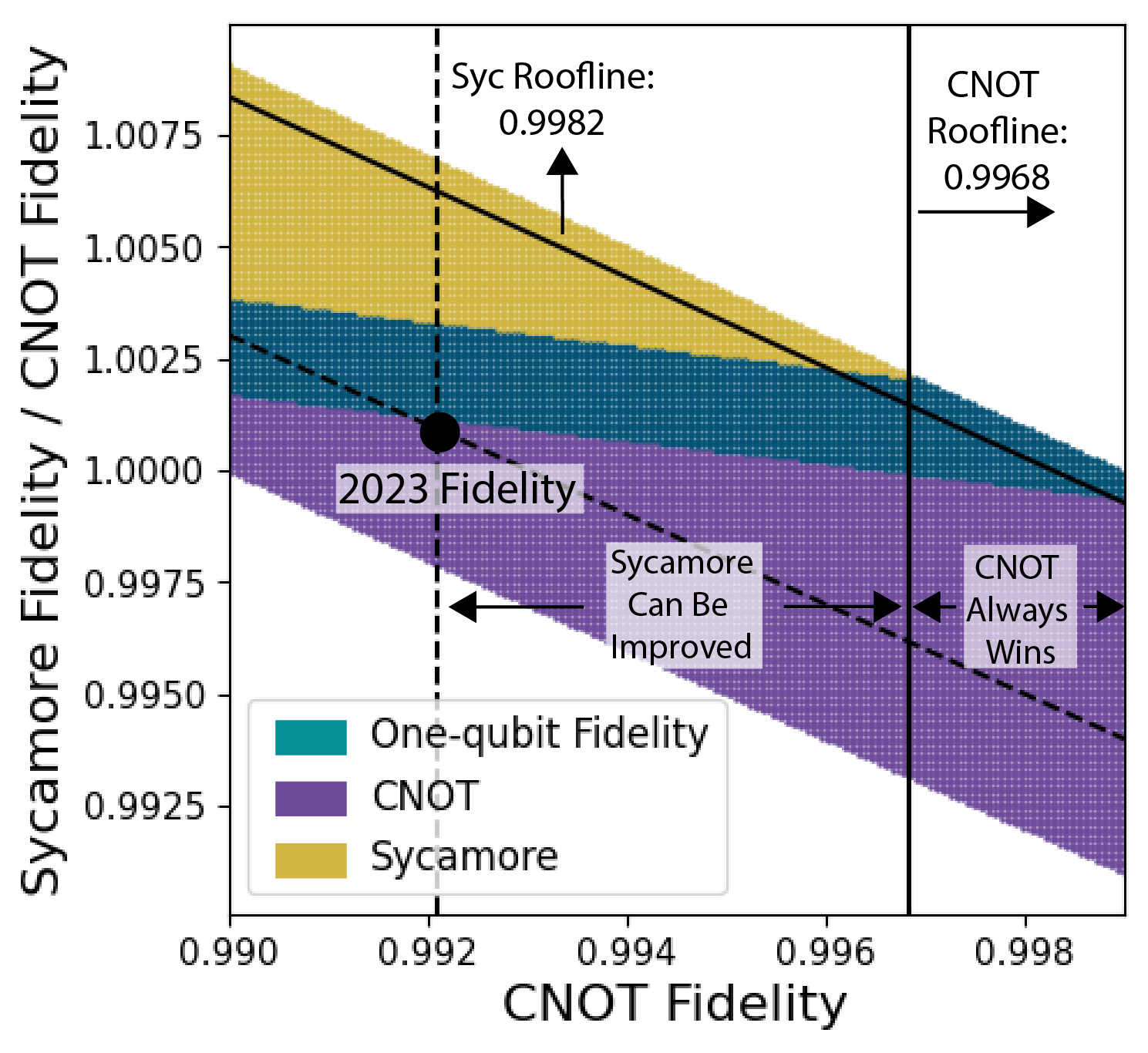}
\end{minipage}%
\hfill
\begin{minipage}{\columnwidth}
    \centering
    \includegraphics[keepaspectratio=true,scale=0.27]{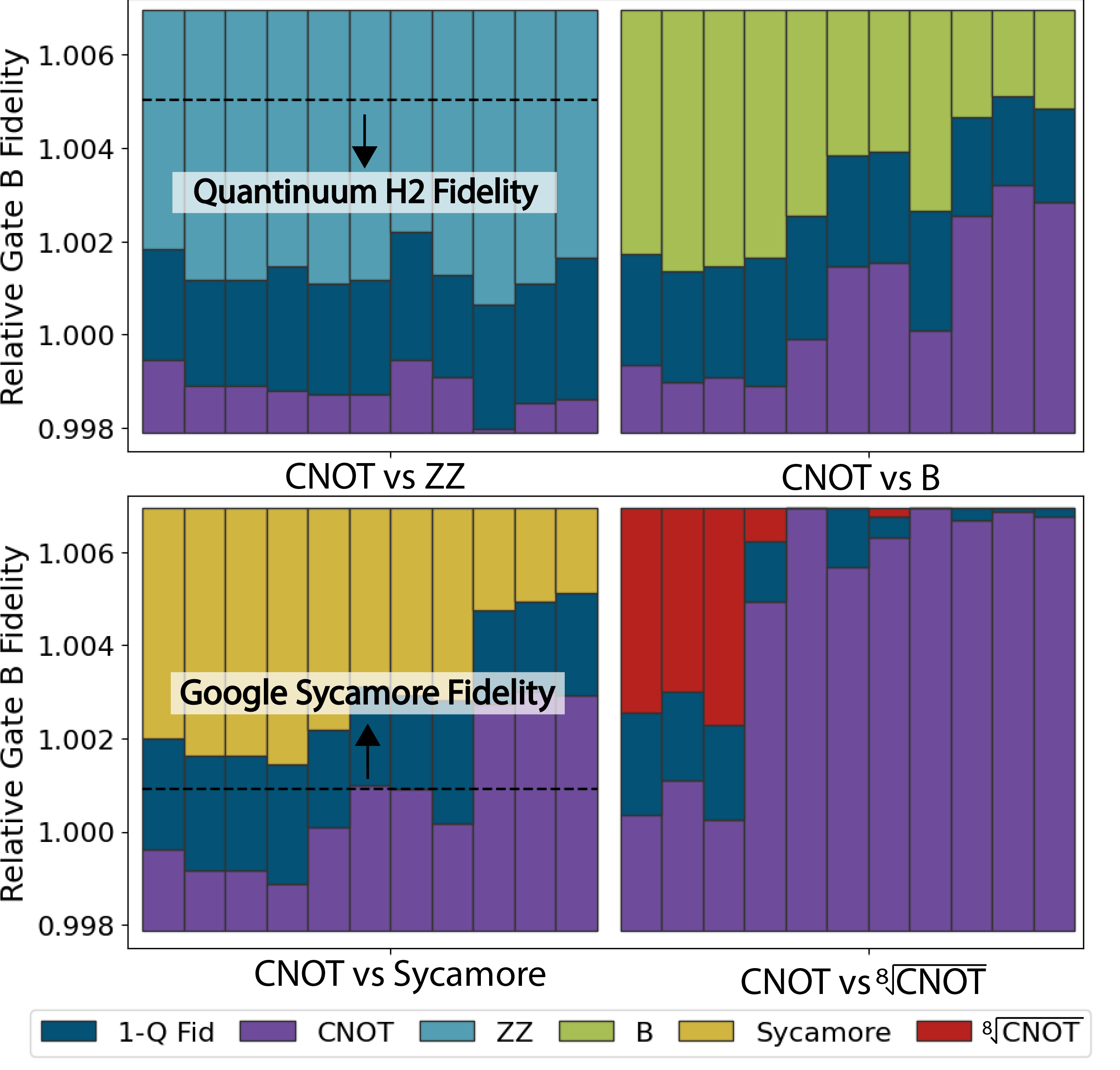}
\end{minipage}
\begin{minipage}{\columnwidth}
    \vspace*{0.14in}
    \caption{\footnotesize \it Machine capability to execute the adder\_9 algorithm. CNOT fidelity is on the x-axis and the relative Sycamore fidelity on the y-axis. We plot the winning machine at each point. The middle area shows where the choice of best machine is a function of the single-qubit fidelity. In the other areas, each machine wins irrespective of 1-qubit gate fidelity. The published 2-qubit gate fidelities are shown with the black dotted lines: interestingly the CNOT based system is better, despite the better Sycamore gate fidelity.}
    \label{fig:adder9_cross}
\end{minipage}%
\hfill
\begin{minipage}{\columnwidth}
    \vspace*{0.1in}
    \caption{\footnotesize \it Gate set comparison against CNOT. CNOT
      fidelity fixed to the IBM Falcon.  Each configuration can be
      improved by tuning the 2-qubit gates. Encouragingly, low entanglement
      gates $\sqrt[8]{CNOT}$ can provide better overall circuit
      fidelity for some algorithms. The bars correspond to the circuit families: TFIM, QAE, QFT, QAOA, QPE, Adders, Grover, Shor, Hubbard, QML, and VQE.}
    \label{fig:y_delt_fig}
\end{minipage}
\end{figure*}

\begin{figure*}
\vspace*{-0.15in}
    \centering
    \includegraphics[keepaspectratio=true,scale=0.31]{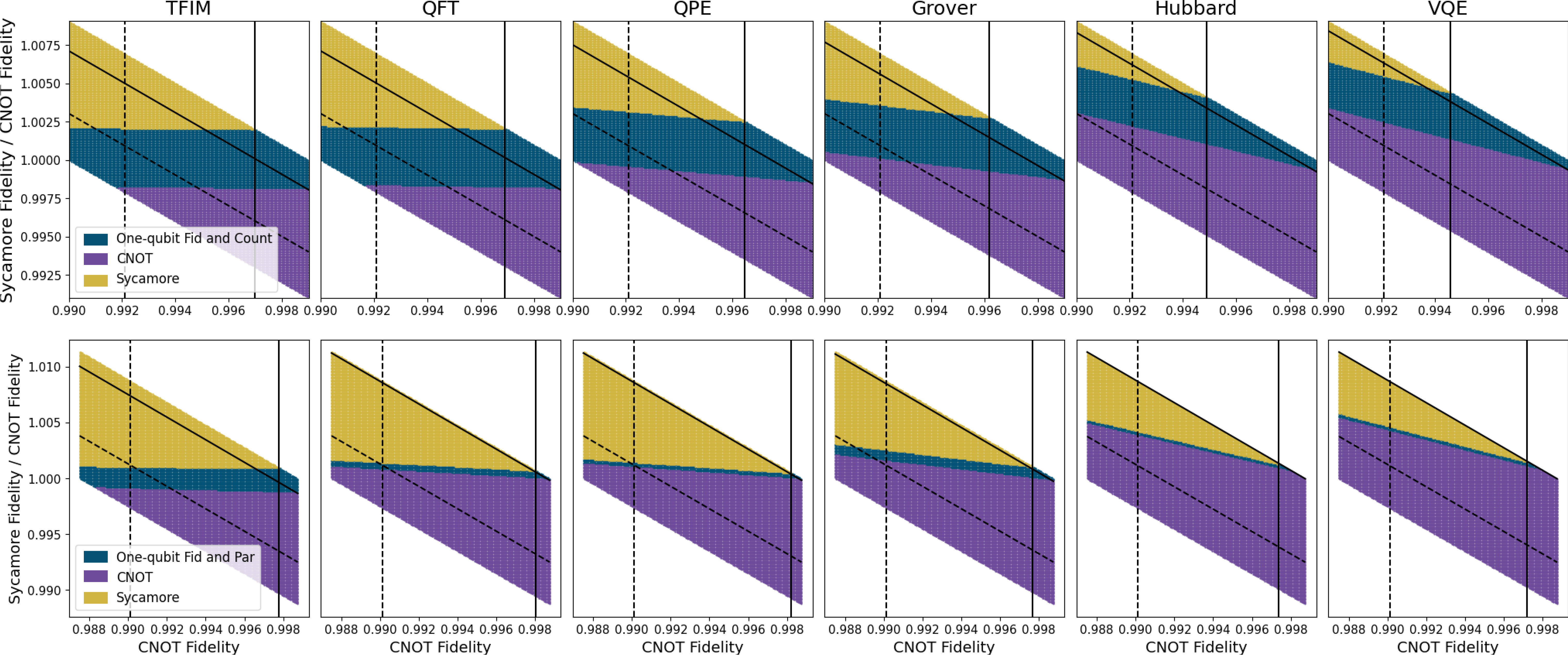}
    \caption{\footnotesize \it Comparison of CNOT and Sycamore based systems when varying 1-qubit gate fidelity. The black dotted lines show the current NISQ fidelities for the two compared machine (IBM Falcon vs. Google Sycamore). Regions labeled Sycamore and CNOT denote which configuration performs best independent of 1-qubit gate fidelity. In the ``One-qubit'' region, behavior is determined by 1-qubit and 2-qubit gate fidelity. }
    \label{fig:all_fids_comp}
\end{figure*}

The results are summarized in Figure~\ref{fig:adder9_cross}.
Each point represents a two-system configuration we are comparing, with the 2-qubit fidelities set according to the {\it x} and {\it y} position. We identify three behavioral regions. In two regions, one configuration wins against the other, {\it no matter the single-qubit gate fidelity of the system}. In these regions 2-qubit gate fidelity and expressivity determine system behavior. In the central region, the behavior depends on the fidelity of single-qubit gates: one machine can be improved relative to the other by tuning their single-qubit gate fidelity. For each system we also compute a {\it 2-qubit gate threshold fidelity}, shown with continuous lines: once that is reached on a system, no improvements\footnote{We vary the fidelities within the constraints of our model.} in the other system's 2-qubit gate fidelity will change the overall ordering. We
also plot the published fidelities of the respective hardware gates
with a dotted line. The distance between actual and  threshold fidelity for a gate indicates a window of opportunity to improve the other system.

We refer to this method of relative comparison as a {\it quantum hardware roofline}, as it allows us to compute bounds on the required improvements for a particular system configuration. For example, In Figure~\ref{fig:adder9_cross}, once the CNOT gate reaches the {\it threshold} fidelity of $0.9968$, no improvements in the Sycamore fidelity will outperform a CNOT based machine.

Figures~\ref{fig:y_delt_fig} and~\ref{fig:all_fids_comp} extend these results across algorithm classes and gatesets.
Figure \ref{fig:y_delt_fig} compares several gates against the CNOT gate whose fidelity is fixed to that of the IBM Falcon system. Again, configurations can be improved by
improving only 2-qubit fidelity, or by improving 1-qubit and 2-qubit gate fidelity together. The exact behavior is algorithm and gate set dependent. Encouragingly, low entanglement gates can provide advantages for some algorithms.

Figure \ref{fig:all_fids_comp} shows
the circuit fidelity comparison plot for the Digital Model and the Cyclic Model. Both the current behavior and the roofline fidelities for both gates vary greatly with the target algorithm class. The trends are similar across models, with the Cyclic Model placing a much smaller emphasis on the single-qubit configuration. This is to be expected since the model penalizes the impact of 2-qubit gates. 

\subsection{Single-Qubit Gate Analysis}
The region where machine performance can be improved by tuning 1-qubit gate fidelity is determined by the algorithm gate count together with the actual gate fidelity. To understand the implications for 1-qubit gate design we consider the closure of the behavior across ``any'' algorithm. Therefore, we consider that the 1-qubit gate count is constrained and related to the two-qubit gate count. Based on our empirical data, we loosely choose the lower bound as $\frac{1}{8}$ of the corresponding 2-qubit gate count. As the upper bound we consider twice the 2-qubit gate count: two consecutive U3 gates can be combined into one, so there is at most two 1-qubit gates in between the 2-qubit gates.

Now, we can allow the 1-qubit gate count to vary in this range and analyze how the objective function behavior changes as we fix single-qubit fidelity ($F_1$) for both machines.

\[\pi = (F_1) ^ {n^A_1} \cdot x^ {4153} - (F_1) ^ {n^B_1} \cdot y^ {5944}  \]

\begin{figure}
    \centering
    \includegraphics[keepaspectratio=true,scale=0.5]{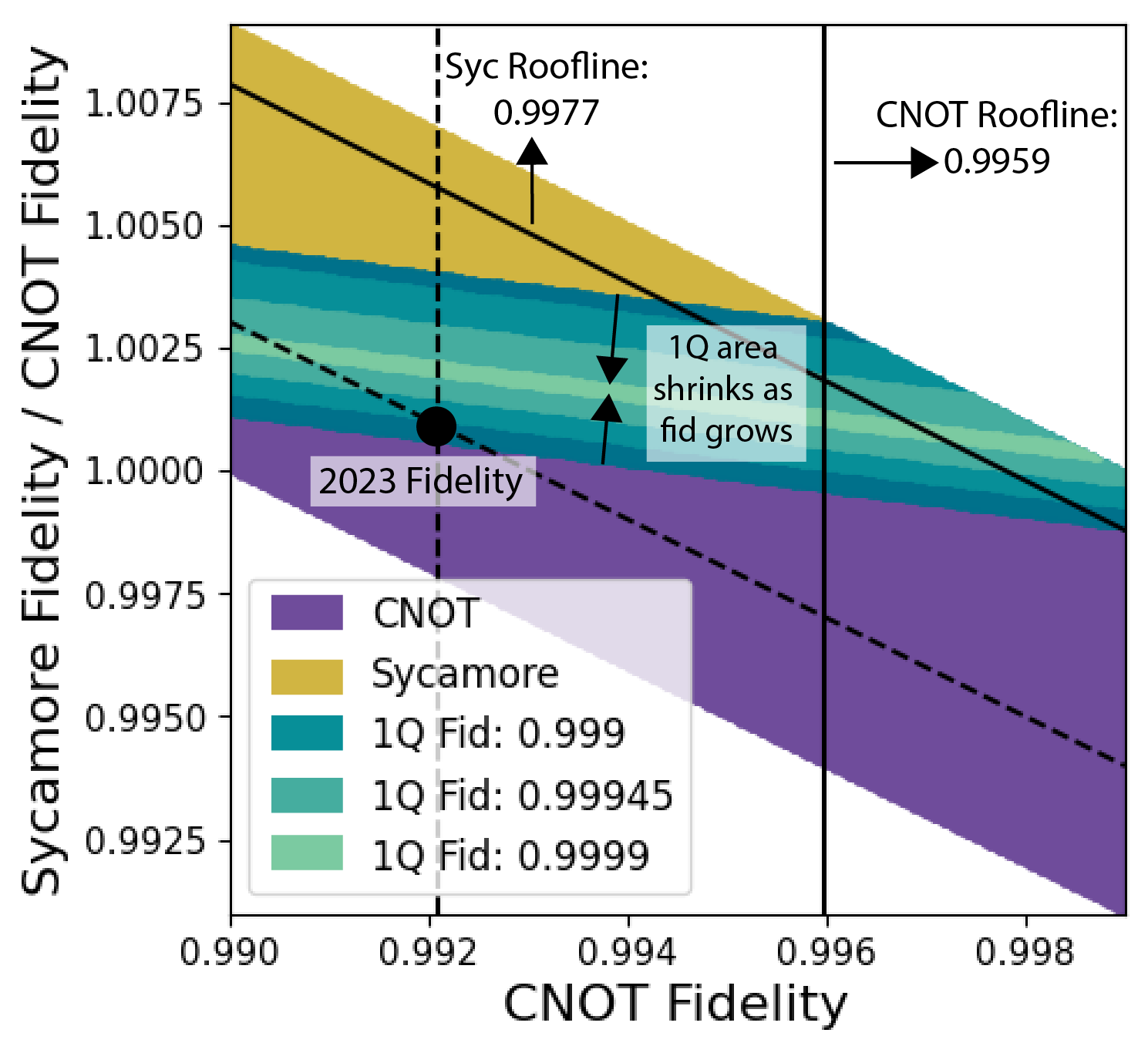}
    \caption{\footnotesize \it This plot shows the behavior of the 1-qubit dependent region as we enforce a higher and higher 1-qubit fidelity. For adder\_9, the region disappears entirely when you reach a 1-qubit fidelity of 0.999988.}
    \label{fig:adder_9_1q}
\end{figure}

The analysis for \emph{adder\_9} on the Falcon and Sycamore machines is shown in Figure \ref{fig:adder_9_1q}.  The 1-qubit dependent region shrinks as the single-qubit fidelity improves, even if we allow for any single-qubit gate count! At some value, the crossover region
completely disappears. We denote this as \textit{single-qubit
  threshold fidelity}: once this is reached, no improvement in 1-qubit gate fidelity will improve comparative performance between two
configurations.

\begin{figure}
    \centering
    \includegraphics[width=.48\textwidth]{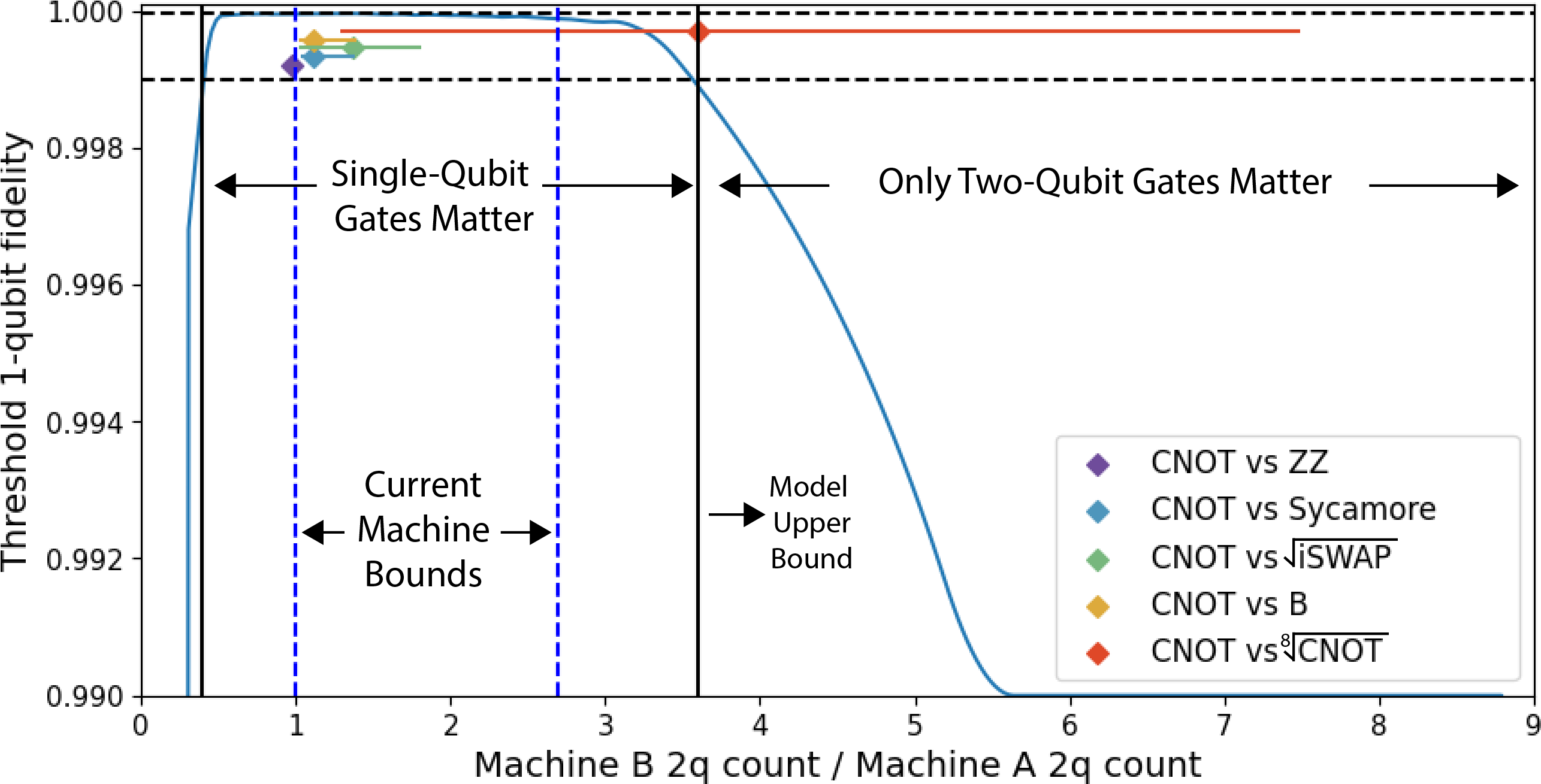}
    \caption{\footnotesize \it The X-axis shows the 2-qubit gate count ratio when comparing the implementations on two machines, while the Y-axis shows the resulting \textit{threshold 1-qubit fidelity}. We also plot the range of 2-qubit gate count ratios we see for each gate compared to CNOT. The black dotted lines show the current NISQ single-qubit fidelity range. When using the upper bound on gate count ratio, the ordering of most machine comparisons is affected by the single-qubit gates. This quickly changes when we consider specific machines as shown in Table \ref{tab:crit_comp}.}
    \label{fig:crit_fid}
\end{figure}

We can solve for the threshold fidelity for different initial 2-qubit gate counts, as present in algorithms. We set up the objective function such that the 2-qubit counts for each machine are related by a ratio, and we vary this ratio along the x-axis in Figure~\ref{fig:crit_fid}. 

\[\pi = y ^ {n^A_1} \cdot {f^{A}_2}^ {n_2^A} - y ^ {n^B_1} \cdot {f^{B}_2}^ {n^B_2 \cdot x}   \]

\begin{figure*}
    \centering
    \includegraphics[width=0.95\textwidth]{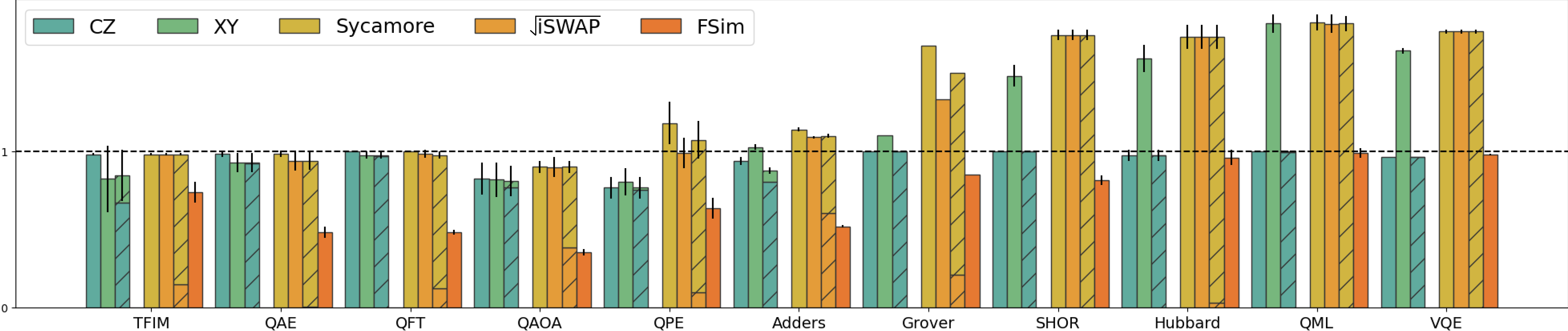}
    \caption{\footnotesize \it 2-qubit gate ratios relative to CNOT, for the Aspen and Google hardware systems. We show homogeneous implementations (CZ, Sycamore, $\sqrt{\text{iSWAP}}$), heterogeneous implementations (CZ+XY, Sycamore+$\sqrt{\text{iSWAP}}$), and parameterized entangling gates (XY, FSim). Heterogeneous implementations (stacked bars) are as good as homogeneous ones. The FSim gate is able to express well all algorithms and it is worth targeting.}
    \label{fig:extra_gates}
\end{figure*}
\begin{figure*}
    \centering
    \includegraphics[width=0.95\textwidth]{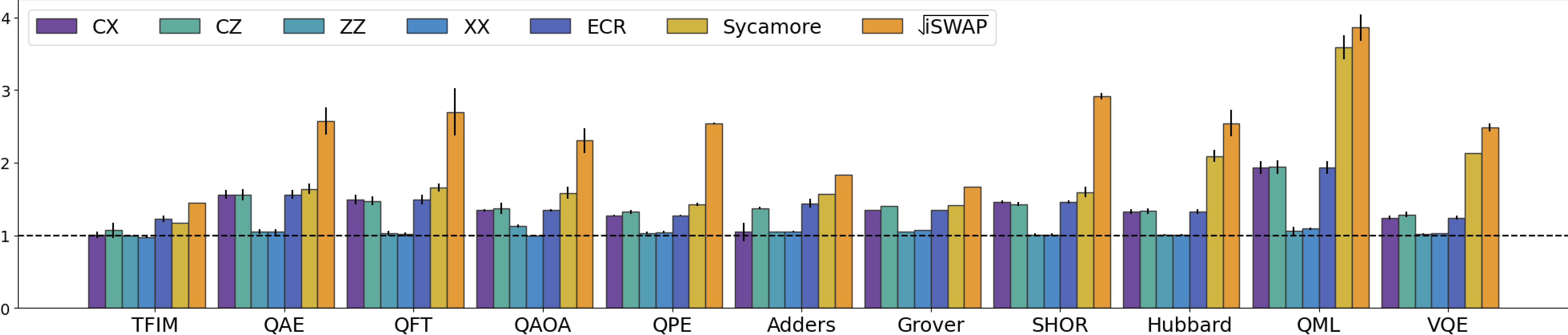}
    \caption{\footnotesize \it 2-qubit gate counts averaged across circuit families mapped to each device topology and gate set.}
    \label{fig:all_fids_topo}
\end{figure*}

As shown in Figure~\ref{fig:crit_fid}, there is an upper bound ratio (3.5) where the threshold fidelity drops below the current worst NISQ-era worst single-qubit fidelities! For any two circuits where the 2-qubit gate ratio is higher than the upper bound (3.5), no improvements in 1-qubit gate on any machine can change relative performance. This explains why for several circuit families, the CNOT machine always beats the $\sqrt[8]{\text{CNOT}}$ machine regardless of the single-qubit configuration!


Using this model, we can derive system-specific upper bound ratios which give direct information about the potential of changing relative performance in practice. Here, we plug into the model the actual 2-qubit gate fidelity ranges of the machines and show results in Table \ref{tab:crit_comp}.  As indicated, the actual upper bounds are around 1.6, much lower than the 3.5 absolute threshold. Overall, this data indicates that for most systems, relative performance orderings can be changed by tuning 1-qubit gates, but for cases such as running QML circuits on Falcon and Sycamore machines, the 1-qubit gates do not matter.

\begin{table}
\vspace*{0.1in}
\centering
\begin{tabular}{|l|l|r|}
\hline
\textbf{Machine 1} & \textbf{Machine 2} & \multicolumn{1}{l|}{\textbf{NISQ Threshold Ratio}} \\ \hline
IBM Falcon         & IBM Eagle          & 1.46 \\ \hline
IBM Falcon         & Google Sycamore    & 1.70  \\ \hline
IBM Falcon         & Quantinuum H2       & 1.06  \\ \hline
IBM Eagle          & Google Sycamore    & 2.69       \\ \hline
IBM Eagle          & Quantinuum H2       & 1.69 \\ \hline
Google Sycamore    & Quantinuum H2        & 1 \\ \hline
\end{tabular}
\caption{\footnotesize \it The upper bound on 2-qubit gates ratio that determines the impact of 1-qubit gate tuning for selected pairs of NISQ machines.  We use gate fidelity provided  by the data-sheets for each machine \cite{Cirq, Qiskit, Quantinuum}. Gate count ratios less than the threshold ratio signify circuits where tuning 1-qubit gates can improve the relative performance. For Sycamore vs. H2 (1), the only way to improve relative performance is by improving 2-qubit gate fidelity, while for the Falcon vs. H2 (1.06) there is a slight window of opportunity.}
\label{tab:crit_comp}
\end{table}

\subsection{Mixing Entangling Gates}
Systems such as Aspen and Sycamore offer multiple entangling gates and parameterized entangling gates. These gates will have different average fidelities and may be used together within a single
circuit. In Figure \ref{fig:extra_gates} we show the 2-qubit gate counts for the Sycamore and Aspen machines when using parameterized and multiple 2-qubit entangling gates within the same
circuit. Heterogeneous gate sets Sycamore+$\sqrt{\text{iSWAP}}$ and
CZ+XY express circuits as well as the Sycamore or CZ gate alone, respectively. Therefore, on these systems choosing the highest fidelity gate for any algorithm may be sufficient.

The parameterized FSim gate leads to significant gate count reduction
when compared to the Sycamore and $\sqrt{\text{iSWAP}}$ gates, while
the XY Gate provided no benefit over the CZ gate. The FSim gate takes two parameters, while the XY gate only has one. This allows the FSim
gate to express complex unitaries more efficiently. Accordingly, FSim implementations can outperform Sycamore gate implementations even with
a larger (0.4\%) drop in fidelity! The circuit quality of the FSim gate may also point to a finite \textit{spanning gate set} of FSim family gates that are able to express circuits as well as the full
parameterized gate. A finite set of constant gates may prove to be easier to calibrate than a fully parameterized gate, leading to a
higher gate fidelity.

\subsection{Topology}
\label{sec:top}
The physical qubit interconnection topology impacts system behavior by: 
\begin{enumerate}
    \item Increasing gate counts when the algorithm logical topology is mismatching, as shown in Figure~\ref{fig:all_fids_topo}.
    \item Adding cross talk due to the device qubit couplings.
\end{enumerate}

\comment{We see that the restrictive grid topology of the superconducting qubits makes a significant difference in the circuit expressivity, and the difference is also a function of the circuit. We see the largest impact for the QML circuits, which on average have increased 2-qubit gate count of 70\%! This makes sense as the two-local circuit inside the QML circuit has a very high logical connectivity. On the other hand the TFIM circuit is only slightly affected as the synthesis algorithm is able to use the available hardware links to express the algorithm. { What is this?}}

\begin{figure}
    \centering
    \includegraphics[width=.45\textwidth]{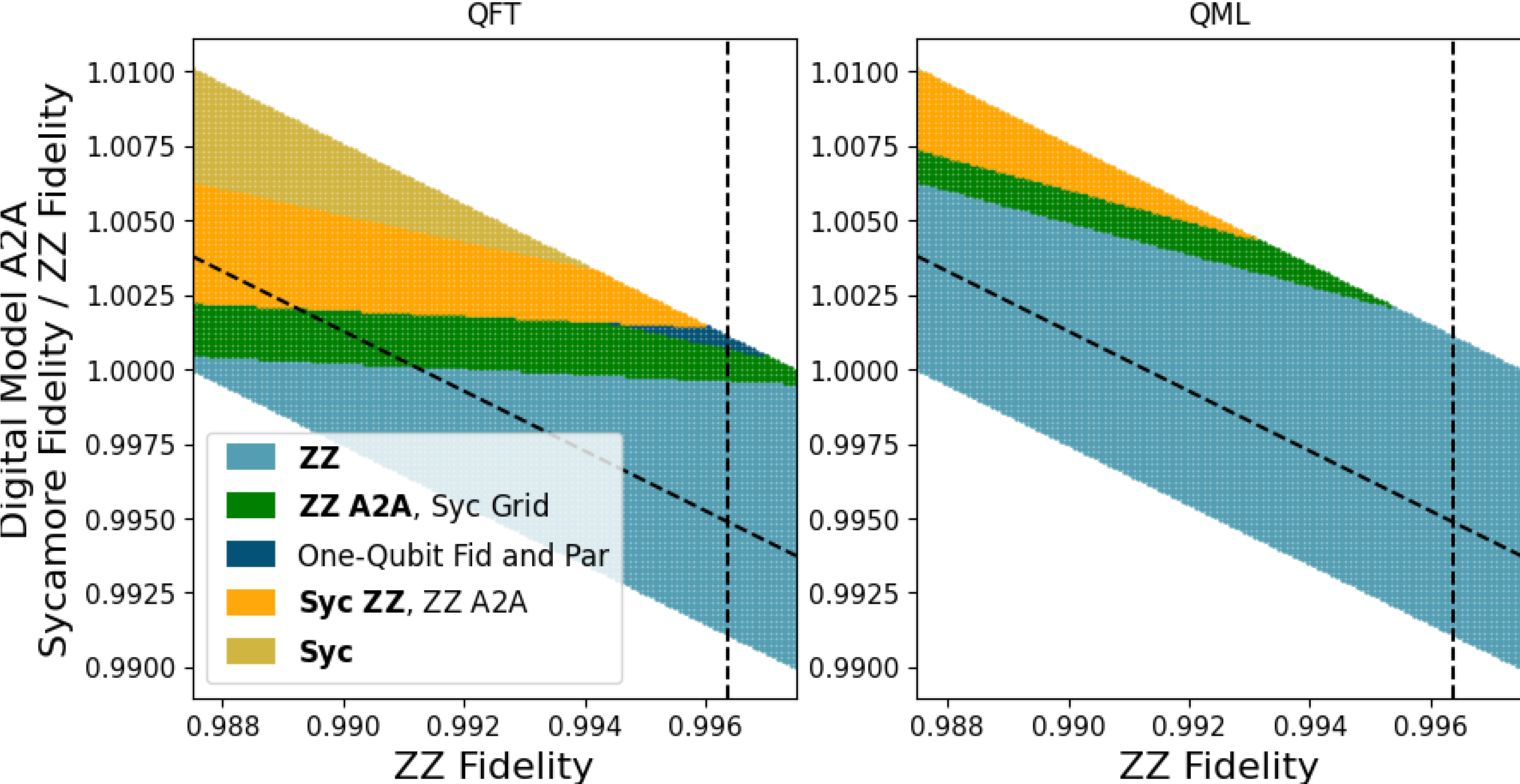}
    \caption{\footnotesize \it Topology effects for (ZZ, Sycamore) X (all-to-all, mesh). Dark blue region - relative performance can be changed by tuning 1-qubit fidelity. Light Blue region - ZZ always performs best, regardless of topology. Yellow region - Sycamore always performs best. In the other regions, the bolded configuration performs best.}
    \label{fig:fid_top0}
\end{figure}

Architects can use our models to answer: "What fidelity improvement is needed to overcome a more restrictive topology?".
In  Figure~\ref{fig:fid_top0} we compare the capabilities of H2 (ZZ, all-to-all) with Sycamore (Sycamore, mesh) using  the cyclic fidelity model. 
As before, we are able to identify ranges where a certain combination (gate, topology) performs best irrespective of 1-qubit gate fidelity, as well as ranges where relative performance depends on 1-qubit gate fidelity as well. The orange region in the graph shows the fidelity range in which Sycamore loses on account of being mapped to a more restrictive topology. The size of this orange region corresponds to the change in ability to express a circuit, and varies by algorithm.

While Figure \ref{fig:fid_top0} is able to model the first effect of topology, we must turn to the coupling-based model introduced in \cite{Erhard_2019} in order to understand cross talk:

    \[\mathbf{F_{c_{\text{top}}}}  = (1 - e_{c1} \cdot C_1)^{n_1} \cdot (1 - e_{c2} \cdot C_2)^{n_2} \] 
\noindent where $C_i$ is the number of other qubits on average that each qubit is coupled to and $e_{ci}$ is the error per coupling for an $i$-qubit gate. $C_i$ is a direct measure of the physical chip topology. For grid topology, usually associated with superconducting qubits, $C_2$ is a constant between 1 and 4. For all-to-all connectivities provided by ion traps $C_2$ is instead $\frac{1}{2}N(N-1)$. The error per coupling is measured by the Cycle Benchmarking protocol, but these numbers are not provided for today's devices.

\begin{figure}
    \centering
    \includegraphics[width=.24\textwidth]{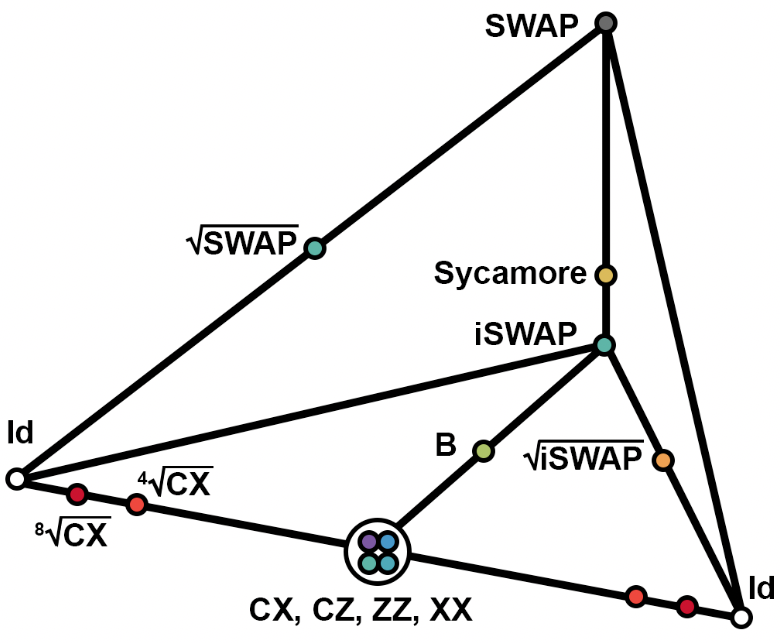}
    \caption{\footnotesize \it Projection of various 2-qubit gates onto the Weyl Chamber. Gates located at the same point represent unitaries that differ by single qubit rotations applied to each qubit (a 1:1 mapping).}
    \label{fig:weyl_chamber}
\end{figure}

\begin{figure}
    \centering
    \begin{subfigure}{0.48\linewidth}
        \includegraphics[width=\linewidth]{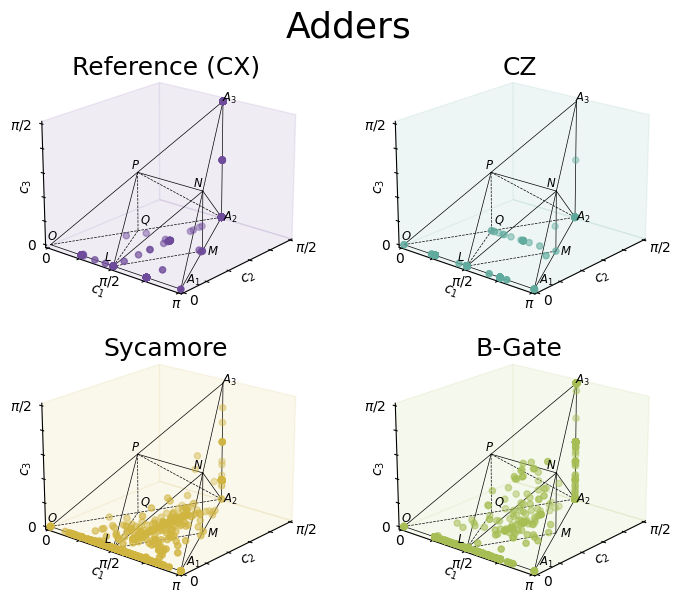}
        \label{fig:group_weyl.a}
    \end{subfigure}
    \begin{subfigure}{0.48\linewidth}
        \includegraphics[width=\linewidth]{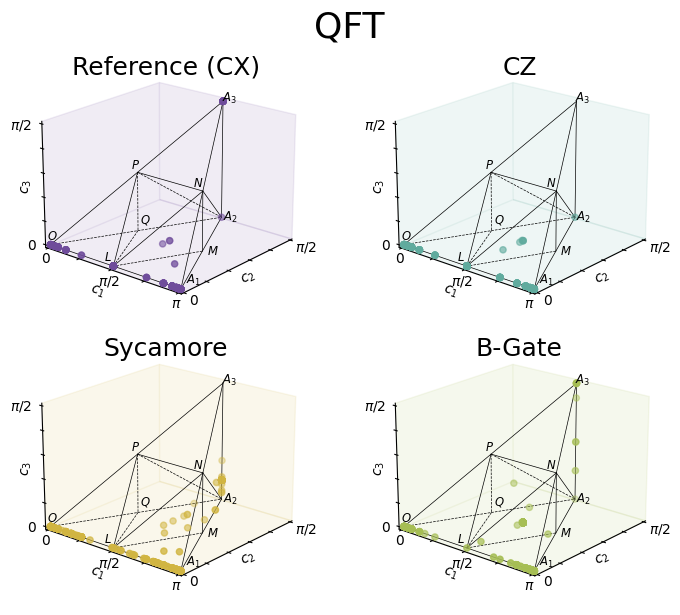}
        \label{fig:group_weyl.b}
    \end{subfigure}
    \vspace*{-0.2in}
    \caption{\footnotesize \it Position in the Weyl chamber of 2-qubit unitaries/blocks that arise in  the Adder group (left) and QFT group (right) for different native gates. The initial spread of Adder unitaries proves to be a very difficult pattern for the Sycamore and B-gates to instantiate, hence the massive increase in gate count. On the other hand, the relatively simple pattern we see in the QFT unitaries (series of controlled rotations) are able to be expressed optimally by Sycamore and B-gates.}
    \label{fig:group_weyl}
\end{figure}

\section{Evaluating Gate Representational Power}
\label{section:discussion}

A gate set's ability to realize a circuit comes down to its
expressivity and entanglement.  Gate \emph{expressivity} identifies a gate's
ability to represent a random two-qubit unitary. Architects often use the
Weyl Chamber to directly visualize gate expressivity \cite{lin_weyl}, as shown in
Figure \ref{fig:weyl_chamber}. The Weyl Chamber removes all local
parameters from a 2-qubit unitary and plots it into a tetrahedron. Most
2-qubit gates can express any 2-qubit unitary in three
applications (along with single qubit rotation gates). The most
expressive gate, the B-gate, can express any unitary in two
applications (it can also easily be realized on a superconducting
machine) \cite{zhang_minimum_2004}.

Gate \emph{entanglement} is a gate's
ability to maximize the entanglement 
between two qubits. CNOTs
and most hardware native gates are maximally entangling, meaning that a single application to two qubits will leave them perfectly
correlated. The $\sqrt[4]{\text{CNOT}}$ and $\sqrt[8]{\text{CNOT}}$
gates trade off entangling power for large
potential gains in gate fidelity.

Several trends have become apparent in our research. Most notably, we see
that with current state-of-the-art compilation, the B-gate does not lead to better circuits, nor does it increase fidelity. This is surprising, because the B-gate is the most expressive 2-qubit gate. On the other hand, we see some equally surprising positive results for the $\sqrt[4]{\text{CNOT}}$ and $\sqrt[8]{\text{CNOT}}$ gates. These low entangling gates require 4 and 8 applications respectively to represent a single CNOT gate. This makes their comparable expressivity in important circuits such as QFT, TFIM, and QAE circuits exciting.

\subsection{Weyl Chamber Distribution of Two-Qubit Algorithm Blocks}

Under our procedure, it is clear that the expressivity and entanglement
of a native 2-qubit gate is not strongly correlated with algorithm
performance under the currently accepted design criteria.  While
expressivity is assessed based on the power to implement random
2-qubit unitaries, optimal implementations of  algorithms impose structure on
the set of 2-qubit unitaries that these gates decompose.

To understand this structure, for any circuit represented in any gate
set, we form maximal 2-qubit unitaries and plot their position
within the Weyl chamber, as shown in Figure~\ref{fig:group_weyl}.

The distribution of 2-qubit unitaries drawn from circuits is not
random. The input structure of the Adder group is spread out (it
contains diverse unitaries), and it so happens that expressive gates,
such as B, have trouble representing some of them. On the other hand,
the input structure of the QFT algorithm is more periodic which allows Sycamore and B-gates to represent it well.

\begin{figure}
    \centering
    \includegraphics[width=0.48\textwidth]{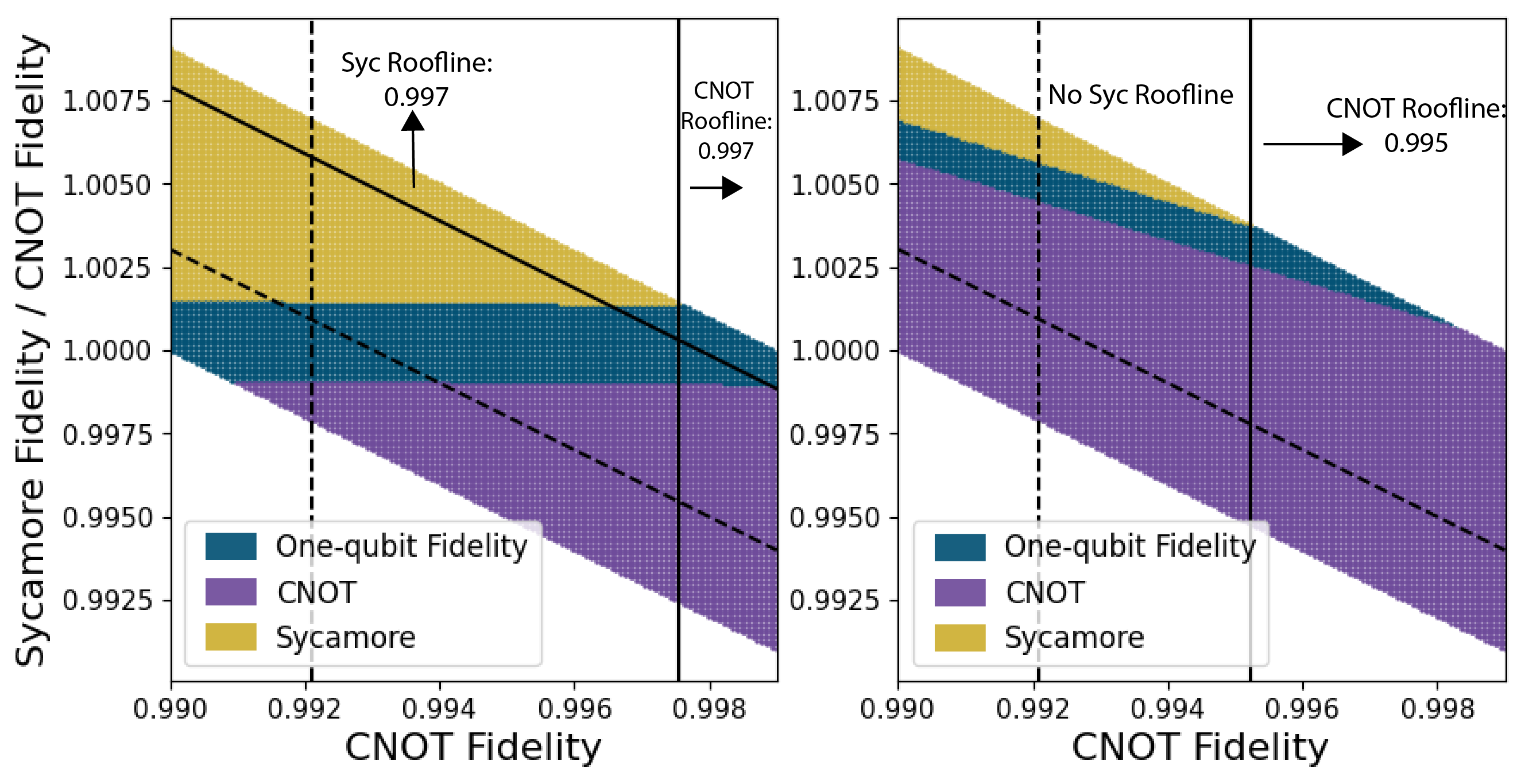}
    \caption{\footnotesize \it Comparison of fidelity plots when compiled with synthesis vs. Cirq. Non-local optimization leads to resource-efficient circuits for different gate sets, which results in vastly different comparisons. Note that under the Cirq compilation, there is no Sycamore roofline.}
    \label{fig:qft_comp_cirq}
\end{figure}

\subsection{Synthesis Derived Gate Selection Criteria}

\label{section:block_dist}

The incorporation of circuit synthesis in our compilation workflow
enables us to derive additional criteria for gate selection and
development. The advantages derived from our flow are due to synthesis'
powerful compilation capabilities.

Given an input circuit, traditional compilers will use local peepholes optimizations,
translating from one gate set to another using analytical, one-to-one gate rewriting
rules for 2-qubit gates. For example, a CNOT is translated into a sequence of two Sycamore gates and additional U3 gates. Any 2-qubit unitary can be represented using at most three CNOT gates. Thus it is expected to use more Sycamore than CNOT gates to represent  an arbitrary 2-qubit gate. As mentioned, it takes at most two B-gates to implement random 2-qubit processes. 

One-to-one gate rewriting has been shown to be less than optimal~\cite{younis_quantum_2022}.
BQSkit's synthesis based compiler~\cite{bqskit} employs a
different strategy. Given an input circuit, BQSKit partitions
it into multi-qubit blocks (partitions). Each partition is optimized
and translated using optimal topology aware direct unitary
synthesis~\cite{davis_qfast}, combined with a powerful synthesis based mapping and routing algorithm~\cite{liu_tackling_2023}. Thus, deploying synthesis
leads to different conclusions than when using vendor provided compilers, as we see in Figure \ref{fig:qft_comp_cirq}.

\begin{figure}
    \centering
    \includegraphics[width=0.35\textwidth]{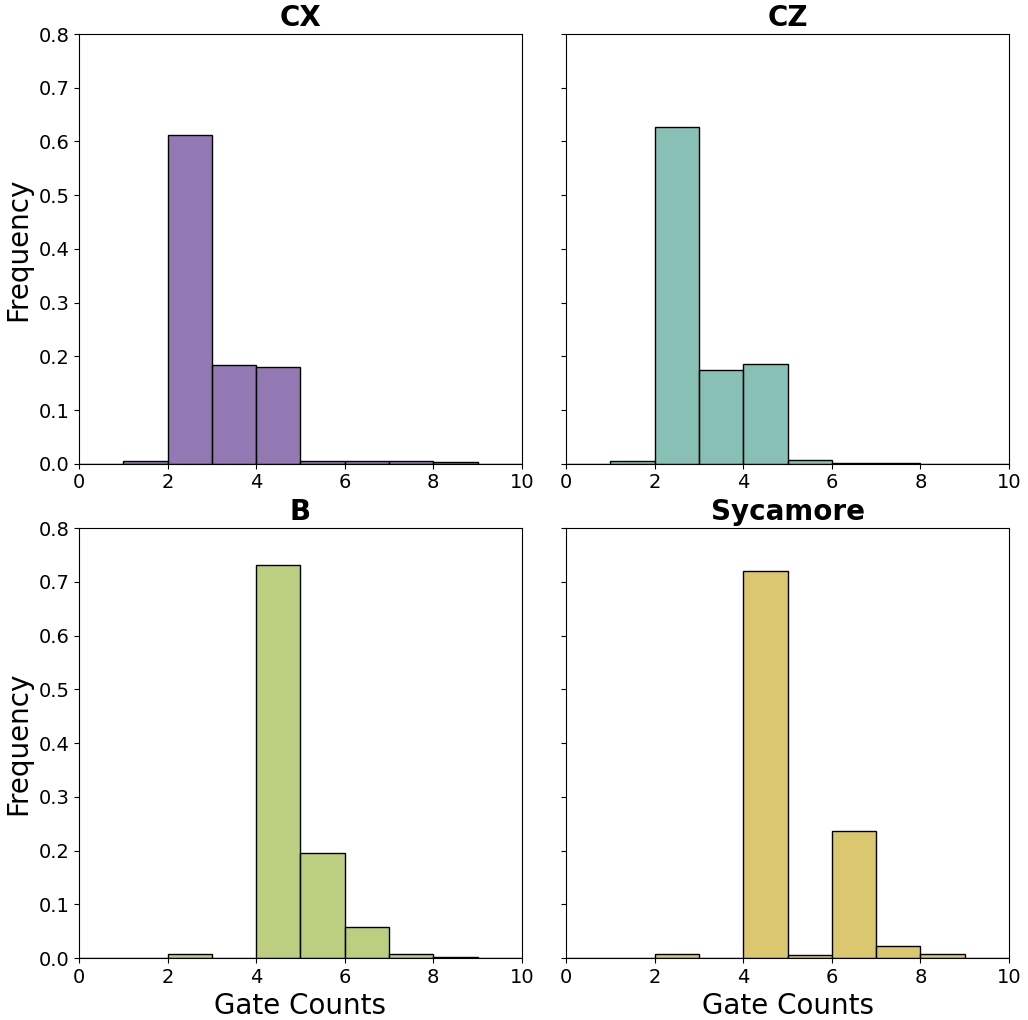}
    \caption{\footnotesize \it Gate counts for 3-qubit unitaries present in the  input and output circuits of the Hubbard group. The input circuit is the CNOT (top left) graph and the output circuits for each 2-qubit gate set is shown. Unsurprisingly CZ and CNOT have similar distributions, while Sycamore leads to a  completely shifted distribution. The B-gate fails to express the simple blocks (fewer than 4 CNOT gates) efficiently but is able to simplify some of the longer blocks.}
    \label{fig:block_distribution}
\end{figure}

More insights can be gained by examining gate representational power
to implement 3-qubit blocks/processes that arise in
algorithm implementations. We use the BQSKit partitioner to decompose
a circuit in maximal 3-qubit gates and then use direct synthesis to generate circuits targeting each
native gate set. This procedure results in implementations for
each block that use an optimal number of 2-qubit gates, irrespective
of gate choice. We plot the gate count distribution of blocks
in Figure \ref{fig:block_distribution}. The CNOT family of gates unsurprisingly have
a one-to-one mapping, but the story changes for
the Sycamore and B-gates. Using Sycamore gates increases gate count, which
is to be expected. On the other hand, we see that the B-gate is able
to better express some more complicated blocks (5-7 CNOT
blocks reduce to 4 B-gates). However, the overall gate count reduction is held back by the B-gate's inability to express simpler blocks efficiently. While a larger block granularity (4+ qubits) would not remove all simple blocks from these circuits, it remains to be seen whether the average increased complexity in each block would allow the B-gate to outperform other gates.

Overall, this analysis indicates that existing gate design criteria should be augmented. In addition to choosing a gate based on attainable fidelity and its representational
power for {\it random} 2-qubit unitaries, the gate representational power for multi-qubit
blocks (e.g. three qubit unitaries), drawn from implementations of real workloads, should be considered.




\section{Discussion}
\label{sec:disc}

While we have introduced and examined several models, we have
emphasized the derivation of the roofline approach for  the digital
fidelity model. A similar derivation can be made for the cyclic model, emphasizing a relative roofline for 1-qubit and 2-qubit process fidelities. Each cycle's fidelity has an absolute parallelism threshold of $\frac{1}{P_i}$ according to the model, and this number will reduce as specific machines/circuits are targeted. We have also emphasized building a roofline model for hardware improvement by considering relative gate fidelity across two configurations. This derivation is useful to hardware designers and algorithm developers. By changing emphasis from fidelity to gate count and circuit depth, a similar derivation can produce roofline models for compiler developers to guide circuit optimization decisions:  ``What mix of gates to choose?; ``Should I reduce gate count or increase gate parallelism?'' etc. 

Multiple studies~ \cite{li_co-design_2021,safi_influence_2023} have
touched on the idea of co-designing quantum hardware, compilers and
algorithms. This paper extends this process by considering
device gate sets that target specific algorithms. For TFIM, QFT, and QAE circuits, we have shown that a designer should maximize gate fidelity even at the cost of expressivity and entanglement
capability. On the other hand, we see highly expressive gates such as the B-gate provide little improvement in overall circuit fidelity.

We see that restricting the topology from an all-to-all connectivity leads to a potentially massive need for higher gate fidelity, varying by algorithm. This means that for Adder-based circuits, Hubbard models, or QML networks, an ion trap  machine with a ZZ or XX gate is best suited.

Our results indicate that unlike classical benchmarking which is
compiler independent, quantum system evaluation and benchmarking is
sensitive to the quality of compilation tools. For the time being, the
compilation workflow requires circuit synthesis for robust inferences.

We believe our methodology will apply beyond the NISQ era into the Fault Tolerant (FT) quantum regime. The model invariant is the circuit gate count. While for NISQ, we directly minimize two-qubit gate count, FT quantum computing requires minimization of error correction overhead, such as non-transversal T gates. Error correction overhead is proportional to the number of one-qubit gates in a circuit and ignores two-qubit gates, leading some circuit decompositions to prioritize single-qubit gate counts~\cite{mottonen2004quantum}. On the other hand, the one-qubit gate count in a circuit is a consequence of the representational power of entangling gates.

\section{Conclusion}
\label{section:conclusion}

In this paper, we introduce a procedure for performing
comparisons between quantum system configurations. In our quantum roofline analysis, we derive bounds on system properties (e.g. gate fidelity) that can be used as a stop criteria for optimization efforts.
We then evaluate machines across a large set of important algorithms
and are able to quantify the trade-off required between gate fidelity,
expressivity, and entanglement for different circuit families in
order to maximize circuit execution fidelity. Our work also shows that
the ability of circuit synthesis to generate resource minimal circuits
is paramount to performance evaluation, and it enables new design
criteria for gate set adoption. We believe our procedure is of interest not only to hardware designers, but compiler and algorithm developers as well.



\bibliographystyle{IEEEtranS}
\bibliography{refs}

\begin{thebibliography}{10}
\providecommand{\url}[1]{#1}
\csname url@samestyle\endcsname
\providecommand{\newblock}{\relax}
\providecommand{\bibinfo}[2]{#2}
\providecommand{\BIBentrySTDinterwordspacing}{\spaceskip=0pt\relax}
\providecommand{\BIBentryALTinterwordstretchfactor}{4}
\providecommand{\BIBentryALTinterwordspacing}{\spaceskip=\fontdimen2\font plus
\BIBentryALTinterwordstretchfactor\fontdimen3\font minus \fontdimen4\font\relax}
\providecommand{\BIBforeignlanguage}[2]{{%
\expandafter\ifx\csname l@#1\endcsname\relax
\typeout{** WARNING: IEEEtranS.bst: No hyphenation pattern has been}%
\typeout{** loaded for the language `#1'. Using the pattern for}%
\typeout{** the default language instead.}%
\else
\language=\csname l@#1\endcsname
\fi
#2}}
\providecommand{\BIBdecl}{\relax}
\BIBdecl

\bibitem{aaronson_complexity-theoretic_2016}
\BIBentryALTinterwordspacing
S.~Aaronson and L.~Chen, ``Complexity-{Theoretic} {Foundations} of {Quantum} {Supremacy} {Experiments},'' \emph{arXiv:1612.05903 [quant-ph]}, Dec. 2016, arXiv: 1612.05903. [Online]. Available: \url{http://arxiv.org/abs/1612.05903}
\BIBentrySTDinterwordspacing

\bibitem{arute_supplementary_2019}
\BIBentryALTinterwordspacing
F.~Arute, K.~Arya, R.~Babbush, D.~Bacon, J.~C. Bardin, R.~Barends, R.~Biswas, S.~Boixo, F.~G. S.~L. Brandao, D.~A. Buell, B.~Burkett, Y.~Chen, Z.~Chen, B.~Chiaro, R.~Collins, W.~Courtney, A.~Dunsworth, E.~Farhi, B.~Foxen, A.~Fowler, C.~Gidney, M.~Giustina, R.~Graff, K.~Guerin, S.~Habegger, M.~P. Harrigan, M.~J. Hartmann, A.~Ho, M.~R. Hoffmann, T.~Huang, T.~S. Humble, S.~V. Isakov, E.~Jeffrey, Z.~Jiang, D.~Kafri, K.~Kechedzhi, J.~Kelly, P.~V. Klimov, S.~Knysh, A.~N. Korotkov, F.~Kostritsa, D.~Landhuis, M.~Lindmark, E.~Lucero, D.~Lyakh, S.~Mandra, J.~R. McClean, M.~McEwen, A.~Megrant, X.~Mi, K.~Michielsen, M.~Mohseni, J.~Mutus, O.~Naaman, M.~Neeley, C.~Neill, M.~Y. Niu, E.~Ostby, A.~Petukhov, J.~C. Platt, C.~Quintana, E.~G. Rieffel, P.~Roushan, N.~C. Rubin, D.~Sank, K.~J. Satzinger, V.~Smelyanskiy, K.~J. Sung, M.~D. Trevithick, A.~Vainsencher, B.~Villalonga, T.~White, Z.~J. Yao, P.~Yeh, A.~Zalcman, H.~Neven, and J.~M. Martinis, ``Supplementary information for "{Quantum} supremacy using a programmable
  superconducting processor",'' \emph{Nature}, vol. 574, no. 7779, pp. 505--510, Oct. 2019, arXiv:1910.11333 [quant-ph]. [Online]. Available: \url{http://arxiv.org/abs/1910.11333}
\BIBentrySTDinterwordspacing

\bibitem{atom_computing_data}
\BIBentryALTinterwordspacing
\emph{Highly Scalable Quantum Computing Wth Atomic Arrays}, Atom Computing, 2023. [Online]. Available: \url{https://atom-computing.com/wp-content/uploads/2022/08/Atom-Computing-Atomic-Arrays.pdf}
\BIBentrySTDinterwordspacing

\bibitem{bao_fluxonium_2022}
\BIBentryALTinterwordspacing
F.~Bao, H.~Deng, D.~Ding, R.~Gao, X.~Gao, C.~Huang, X.~Jiang, H.-S. Ku, Z.~Li, X.~Ma, X.~Ni, J.~Qin, Z.~Song, H.~Sun, C.~Tang, T.~Wang, F.~Wu, T.~Xia, W.~Yu, F.~Zhang, G.~Zhang, X.~Zhang, J.~Zhou, X.~Zhu, Y.~Shi, J.~Chen, H.-H. Zhao, and C.~Deng, ``Fluxonium: an alternative qubit platform for high-fidelity operations,'' \emph{Physical Review Letters}, vol. 129, no.~1, p. 010502, Jun. 2022, arXiv:2111.13504 [quant-ph]. [Online]. Available: \url{http://arxiv.org/abs/2111.13504}
\BIBentrySTDinterwordspacing

\bibitem{barends_transmon}
\BIBentryALTinterwordspacing
R.~Barends, J.~Kelly, A.~Megrant, D.~Sank, E.~Jeffrey, Y.~Chen, Y.~Yin, B.~Chiaro, J.~Mutus, C.~Neill, P.~O'Malley, P.~Roushan, J.~Wenner, T.~C. White, A.~N. Cleland, and J.~M. Martinis, ``Coherent josephson qubit suitable for scalable quantum integrated circuits,'' \emph{Phys. Rev. Lett.}, vol. 111, p. 080502, Aug 2013. [Online]. Available: \url{https://link.aps.org/doi/10.1103/PhysRevLett.111.080502}
\BIBentrySTDinterwordspacing

\bibitem{beauregard_circuit_2003}
\BIBentryALTinterwordspacing
S.~Beauregard, ``\BIBforeignlanguage{en}{Circuit for {Shor}'s algorithm using 2n+3 qubits},'' Feb. 2003, arXiv:quant-ph/0205095. [Online]. Available: \url{http://arxiv.org/abs/quant-ph/0205095}
\BIBentrySTDinterwordspacing

\bibitem{bravyi_fermionic_2002}
\BIBentryALTinterwordspacing
S.~Bravyi and A.~Kitaev, ``Fermionic quantum computation,'' \emph{Annals of Physics}, vol. 298, no.~1, pp. 210--226, May 2002, arXiv:quant-ph/0003137. [Online]. Available: \url{http://arxiv.org/abs/quant-ph/0003137}
\BIBentrySTDinterwordspacing

\bibitem{cerezo_challenges_2022}
\BIBentryALTinterwordspacing
M.~Cerezo, G.~Verdon, H.-Y. Huang, L.~Cincio, and P.~J. Coles, ``\BIBforeignlanguage{en}{Challenges and opportunities in quantum machine learning},'' \emph{\BIBforeignlanguage{en}{Nature Computational Science}}, vol.~2, no.~9, pp. 567--576, Sep. 2022, number: 9 Publisher: Nature Publishing Group. [Online]. Available: \url{https://www.nature.com/articles/s43588-022-00311-3}
\BIBentrySTDinterwordspacing

\bibitem{chen_benchmarking_2023}
\BIBentryALTinterwordspacing
J.-S. Chen, E.~Nielsen, M.~Ebert, V.~Inlek, K.~Wright, V.~Chaplin, A.~Maksymov, E.~Páez, A.~Poudel, P.~Maunz, and J.~Gamble, ``Benchmarking a trapped-ion quantum computer with 29 algorithmic qubits,'' Aug. 2023, arXiv:2308.05071 [quant-ph]. [Online]. Available: \url{http://arxiv.org/abs/2308.05071}
\BIBentrySTDinterwordspacing

\bibitem{cirac_ions}
\BIBentryALTinterwordspacing
J.~I. Cirac and P.~Zoller, ``Quantum computations with cold trapped ions,'' \emph{Phys. Rev. Lett.}, vol.~74, pp. 4091--4094, May 1995. [Online]. Available: \url{https://link.aps.org/doi/10.1103/PhysRevLett.74.4091}
\BIBentrySTDinterwordspacing

\bibitem{Cowtan_2020}
\BIBentryALTinterwordspacing
A.~Cowtan, S.~Dilkes, R.~Duncan, W.~Simmons, and S.~Sivarajah, ``Phase gadget synthesis for shallow circuits,'' \emph{Electronic Proceedings in Theoretical Computer Science}, vol. 318, p. 213–228, May 2020. [Online]. Available: \url{http://dx.doi.org/10.4204/EPTCS.318.13}
\BIBentrySTDinterwordspacing

\bibitem{cross_validating_2019}
\BIBentryALTinterwordspacing
A.~W. Cross, L.~S. Bishop, S.~Sheldon, P.~D. Nation, and J.~M. Gambetta, ``Validating quantum computers using randomized model circuits,'' \emph{Physical Review A}, vol. 100, no.~3, p. 032328, Sep. 2019, arXiv:1811.12926 [quant-ph]. [Online]. Available: \url{http://arxiv.org/abs/1811.12926}
\BIBentrySTDinterwordspacing

\bibitem{davis_qfast}
M.~G. Davis, E.~Smith, A.~Tudor, K.~Sen, I.~Siddiqi, and C.~Iancu, ``Towards optimal topology aware quantum circuit synthesis,'' in \emph{2020 IEEE International Conference on Quantum Computing and Engineering (QCE)}, 2020, pp. 223--234.

\bibitem{Cirq}
\BIBentryALTinterwordspacing
C.~Developers, ``Cirq,'' Jul. 2023. [Online]. Available: \url{https://doi.org/10.5281/zenodo.8161252}
\BIBentrySTDinterwordspacing

\bibitem{Erhard_2019}
\BIBentryALTinterwordspacing
A.~Erhard, J.~J. Wallman, L.~Postler, M.~Meth, R.~Stricker, E.~A. Martinez, P.~Schindler, T.~Monz, J.~Emerson, and R.~Blatt, ``Characterizing large-scale quantum computers via cycle benchmarking,'' \emph{Nature Communications}, vol.~10, no.~1, nov 2019. [Online]. Available: \url{https://doi.org/10.1038%2Fs41467-019-13068-7}
\BIBentrySTDinterwordspacing

\bibitem{farhi_quantum_2014}
\BIBentryALTinterwordspacing
E.~Farhi, J.~Goldstone, and S.~Gutmann, ``A {Quantum} {Approximate} {Optimization} {Algorithm},'' Nov. 2014, arXiv:1411.4028 [quant-ph]. [Online]. Available: \url{http://arxiv.org/abs/1411.4028}
\BIBentrySTDinterwordspacing

\bibitem{herman_survey_2022}
\BIBentryALTinterwordspacing
D.~Herman, C.~Googin, X.~Liu, A.~Galda, I.~Safro, Y.~Sun, M.~Pistoia, and Y.~Alexeev, ``\BIBforeignlanguage{en}{A {Survey} of {Quantum} {Computing} for {Finance}},'' Jun. 2022, arXiv:2201.02773 [quant-ph, q-fin]. [Online]. Available: \url{http://arxiv.org/abs/2201.02773}
\BIBentrySTDinterwordspacing

\bibitem{horodecki_general_1999}
\BIBentryALTinterwordspacing
P.~Horodecki, M.~Horodecki, and R.~Horodecki, ``General teleportation channel, singlet fraction and quasi-distillation,'' Mar. 1999, arXiv:quant-ph/9807091. [Online]. Available: \url{http://arxiv.org/abs/quant-ph/9807091}
\BIBentrySTDinterwordspacing

\bibitem{ionq_data}
\BIBentryALTinterwordspacing
{IonQ Staff}, ``\BIBforeignlanguage{en}{{IonQ} {Forte}: {The} {First} {Software}-{Configurable} {Quantum} {Computer}}.'' [Online]. Available: \url{https://ionq.com/resources/ionq-forte-first-configurable-quantum-computer}
\BIBentrySTDinterwordspacing

\bibitem{Jaksch_neutral}
\BIBentryALTinterwordspacing
D.~Jaksch, J.~I. Cirac, P.~Zoller, S.~L. Rolston, R.~C\^ot\'e, and M.~D. Lukin, ``Fast quantum gates for neutral atoms,'' \emph{Phys. Rev. Lett.}, vol.~85, pp. 2208--2211, Sep 2000. [Online]. Available: \url{https://link.aps.org/doi/10.1103/PhysRevLett.85.2208}
\BIBentrySTDinterwordspacing

\bibitem{kim_evidence_2023}
\BIBentryALTinterwordspacing
Y.~Kim, A.~Eddins, S.~Anand, K.~X. Wei, E.~van~den Berg, S.~Rosenblatt, H.~Nayfeh, Y.~Wu, M.~Zaletel, K.~Temme, and A.~Kandala, ``\BIBforeignlanguage{en}{Evidence for the utility of quantum computing before fault tolerance},'' \emph{\BIBforeignlanguage{en}{Nature}}, vol. 618, no. 7965, pp. 500--505, Jun. 2023, number: 7965 Publisher: Nature Publishing Group. [Online]. Available: \url{https://www.nature.com/articles/s41586-023-06096-3}
\BIBentrySTDinterwordspacing

\bibitem{knill_randomized_2008}
\BIBentryALTinterwordspacing
E.~Knill, D.~Leibfried, R.~Reichle, J.~Britton, R.~B. Blakestad, J.~D. Jost, C.~Langer, R.~Ozeri, S.~Seidelin, and D.~J. Wineland, ``Randomized {Benchmarking} of {Quantum} {Gates},'' \emph{Physical Review A}, vol.~77, no.~1, p. 012307, Jan. 2008, arXiv:0707.0963 [quant-ph]. [Online]. Available: \url{http://arxiv.org/abs/0707.0963}
\BIBentrySTDinterwordspacing

\bibitem{li2019tackling}
G.~Li, Y.~Ding, and Y.~Xie, ``Tackling the qubit mapping problem for nisq-era quantum devices,'' 2019.

\bibitem{li_co-design_2021}
\BIBentryALTinterwordspacing
G.~Li, A.~Wu, Y.~Shi, A.~Javadi-Abhari, Y.~Ding, and Y.~Xie, ``On the {Co}-{Design} of {Quantum} {Software} and {Hardware},'' in \emph{Proceedings of the {Eight} {Annual} {ACM} {International} {Conference} on {Nanoscale} {Computing} and {Communication}}, ser. {NANOCOM} '21.\hskip 1em plus 0.5em minus 0.4em\relax New York, NY, USA: Association for Computing Machinery, Sep. 2021, pp. 1--7. [Online]. Available: \url{https://dl.acm.org/doi/10.1145/3477206.3477464}
\BIBentrySTDinterwordspacing

\bibitem{lin_weyl}
S.~F. Lin, S.~Sussman, C.~Duckering, P.~S. Mundada, J.~M. Baker, R.~S. Kumar, A.~A. Houck, and F.~T. Chong, ``Let each quantum bit choose its basis gates,'' in \emph{2022 55th IEEE/ACM International Symposium on Microarchitecture (MICRO)}, 2022, pp. 1042--1058.

\bibitem{liu_tackling_2023}
\BIBentryALTinterwordspacing
J.~Liu, E.~Younis, M.~Weiden, P.~Hovland, J.~Kubiatowicz, and C.~Iancu, ``Tackling the {Qubit} {Mapping} {Problem} with {Permutation}-{Aware} {Synthesis},'' May 2023, arXiv:2305.02939 [quant-ph]. [Online]. Available: \url{http://arxiv.org/abs/2305.02939}
\BIBentrySTDinterwordspacing

\bibitem{magesan_robust_2011}
\BIBentryALTinterwordspacing
E.~Magesan, J.~M. Gambetta, and J.~Emerson, ``Robust randomized benchmarking of quantum processes,'' \emph{Physical Review Letters}, vol. 106, no.~18, p. 180504, May 2011, arXiv:1009.3639 [quant-ph]. [Online]. Available: \url{http://arxiv.org/abs/1009.3639}
\BIBentrySTDinterwordspacing

\bibitem{magesan_characterizing_2012}
\BIBentryALTinterwordspacing
E.~Magesan, J.~M. Gambetta, and J.~Emerson, ``Characterizing {Quantum} {Gates} via {Randomized} {Benchmarking},'' \emph{Physical Review A}, vol.~85, no.~4, p. 042311, Apr. 2012, arXiv:1109.6887 [quant-ph]. [Online]. Available: \url{http://arxiv.org/abs/1109.6887}
\BIBentrySTDinterwordspacing

\bibitem{mandviwalla_implementing_2018}
\BIBentryALTinterwordspacing
A.~Mandviwalla, K.~Ohshiro, and B.~Ji, ``Implementing {Grover}’s {Algorithm} on the {IBM} {Quantum} {Computers},'' in \emph{2018 {IEEE} {International} {Conference} on {Big} {Data} ({Big} {Data})}, Dec. 2018, pp. 2531--2537. [Online]. Available: \url{https://ieeexplore.ieee.org/document/8622457}
\BIBentrySTDinterwordspacing

\bibitem{openfermion}
\BIBentryALTinterwordspacing
J.~R. McClean, I.~D. Kivlichan, D.~S. Steiger, Y.~Cao, E.~S. Fried, C.~Gidney, T.~Häner, V.~Havlíček, Z.~Jiang, M.~Neeley, J.~Romero, N.~Rubin, N.~P.~D. Sawaya, K.~Setia, S.~Sim, W.~Sun, K.~Sung, and R.~Babbush, ``Openfermion: The electronic structure package for quantum computers,'' 2017, cite arxiv:1710.07629. [Online]. Available: \url{http://arxiv.org/abs/1710.07629}
\BIBentrySTDinterwordspacing

\bibitem{mottonen2004quantum}
M.~M{\"o}tt{\"o}nen, J.~J. Vartiainen, V.~Bergholm, and M.~M. Salomaa, ``Quantum circuits for general multiqubit gates,'' \emph{Physical review letters}, vol.~93, no.~13, p. 130502, 2004.

\bibitem{Nielsen_2002}
\BIBentryALTinterwordspacing
M.~A. Nielsen, ``A simple formula for the average gate fidelity of a quantum dynamical operation,'' \emph{Physics Letters A}, vol. 303, no.~4, pp. 249--252, oct 2002. [Online]. Available: \url{https://doi.org/10.1016%2Fs0375-9601%2802%2901272-0}
\BIBentrySTDinterwordspacing

\bibitem{Bassman_Oftelie_2022}
\BIBentryALTinterwordspacing
L.~B. Oftelie, R.~V. Beeumen, E.~Younis, E.~Smith, C.~Iancu, and W.~A. de~Jong, ``Constant-depth circuits for dynamic simulations of materials on quantum computers,'' \emph{Materials Theory}, vol.~6, no.~1, mar 2022. [Online]. Available: \url{https://doi.org/10.1186%2Fs41313-022-00043-x}
\BIBentrySTDinterwordspacing

\bibitem{peruzzo_variational_2014}
\BIBentryALTinterwordspacing
A.~Peruzzo, J.~McClean, P.~Shadbolt, M.-H. Yung, X.-Q. Zhou, P.~J. Love, A.~Aspuru-Guzik, and J.~L. O'Brien, ``A variational eigenvalue solver on a quantum processor,'' \emph{Nature Communications}, vol.~5, no.~1, p. 4213, Jul. 2014, arXiv:1304.3061 [physics, physics:quant-ph]. [Online]. Available: \url{http://arxiv.org/abs/1304.3061}
\BIBentrySTDinterwordspacing

\bibitem{Qiskit}
{Qiskit contributors}, ``Qiskit: An open-source framework for quantum computing,'' 2023.

\bibitem{Quantinuum}
\emph{Quantinuum System Model H2}, Quantinuum, 2023.

\bibitem{squander}
P.~Rakyta and Z.~Zimborás, ``Efficient quantum gate decomposition via adaptive circuit compression,'' 2022.

\bibitem{rigetti_data}
\emph{Aspen-M-3 Quantum Processor}, Rigetti {QCS}, 2023.

\bibitem{safi_influence_2023}
\BIBentryALTinterwordspacing
H.~Safi, K.~Wintersperger, and W.~Mauerer, ``Influence of {HW}-{SW}-{Co}-{Design} on {Quantum} {Computing} {Scalability},'' Jun. 2023, arXiv:2306.04246 [quant-ph]. [Online]. Available: \url{http://arxiv.org/abs/2306.04246}
\BIBentrySTDinterwordspacing

\bibitem{shin_phonon-driven_2018}
\BIBentryALTinterwordspacing
D.~Shin, H.~Hübener, U.~De~Giovannini, H.~Jin, A.~Rubio, and N.~Park, ``\BIBforeignlanguage{en}{Phonon-driven spin-{Floquet} magneto-valleytronics in {MoS2}},'' \emph{\BIBforeignlanguage{en}{Nature Communications}}, vol.~9, no.~1, p. 638, Feb. 2018, number: 1 Publisher: Nature Publishing Group. [Online]. Available: \url{https://www.nature.com/articles/s41467-018-02918-5}
\BIBentrySTDinterwordspacing

\bibitem{sivarajah_tketrangle_2021}
\BIBentryALTinterwordspacing
S.~Sivarajah, S.~Dilkes, A.~Cowtan, W.~Simmons, A.~Edgington, and R.~Duncan, ``t\${\textbar}\$ket\${\textbackslash}rangle\$ : {A} {Retargetable} {Compiler} for {NISQ} {Devices},'' \emph{Quantum Science and Technology}, vol.~6, no.~1, p. 014003, Jan. 2021, arXiv:2003.10611 [quant-ph]. [Online]. Available: \url{http://arxiv.org/abs/2003.10611}
\BIBentrySTDinterwordspacing

\bibitem{tucci2005introduction}
R.~R. Tucci, ``An introduction to cartan's kak decomposition for qc programmers,'' 2005.

\bibitem{qseed}
\BIBentryALTinterwordspacing
M.~Weiden, E.~Younis, J.~Kalloor, J.~Kubiatowicz, and C.~Iancu, ``Improving {Quantum} {Circuit} {Synthesis} with {Machine} {Learning},'' Jun. 2023, arXiv:2306.05622 [quant-ph]. [Online]. Available: \url{http://arxiv.org/abs/2306.05622}
\BIBentrySTDinterwordspacing

\bibitem{younis_quantum_2022}
\BIBentryALTinterwordspacing
E.~Younis and C.~Iancu, ``Quantum {Circuit} {Optimization} and {Transpilation} via {Parameterized} {Circuit} {Instantiation},'' Jun. 2022, arXiv:2206.07885 [quant-ph]. [Online]. Available: \url{http://arxiv.org/abs/2206.07885}
\BIBentrySTDinterwordspacing

\bibitem{bqskit}
\BIBentryALTinterwordspacing
E.~Younis, C.~C. Iancu, W.~Lavrijsen, M.~Davis, E.~Smith, and USDOE, ``Berkeley quantum synthesis toolkit (bqskit) v1,'' 4 2021. [Online]. Available: \url{https://www.osti.gov//servlets/purl/1785933}
\BIBentrySTDinterwordspacing

\bibitem{zhang_minimum_2004}
\BIBentryALTinterwordspacing
J.~Zhang, J.~Vala, S.~Sastry, and K.~B. Whaley, ``Minimum construction of two-qubit quantum operations,'' \emph{Physical Review Letters}, vol.~93, no.~2, p. 020502, Jul. 2004, arXiv:quant-ph/0312193. [Online]. Available: \url{http://arxiv.org/abs/quant-ph/0312193}
\BIBentrySTDinterwordspacing

\end{thebibliography}

\end{document}